\documentclass[prx,aps,amsart,amssymb,epsfig,showpacs,twocolumn,eqsecnum,superscriptaddress]{revtex4-1}

\usepackage{amsmath,cleveref,amssymb,amsfonts,graphicx,multirow,color,bm,textcomp,mathtools}
\usepackage{paralist}
\usepackage{srcltx}
\usepackage[all,cmtip]{xy}
\usepackage{mathrsfs}
\usepackage{srcltx,soul,subfigure}
\usepackage{threeparttable,dsfont,bbm}

\usepackage{hyperref,cleveref}

\crefname{equation}{Eq.}{Eqs.}
\newcommand{\vk}{{\bf k}}

\newcommand{\bee}{\begin{eqnarray}}
\newcommand{\ee}{\end{eqnarray}}
\newcommand{\bma}{\begin{pmatrix}}
\newcommand{\ema}{\end{pmatrix}}
\newcommand{\balig}{\begin{align}}
\newcommand{\ealig}{\end{align}}

\newcommand{\ba}{\begin{align}}
\newcommand{\ea}{\end{align}}

\newcommand{\ignore}[1]{}

\newcommand{\Csix}{\tilde{C}_{6,p}}

\begin{document}

\title{Topological band crossings in hexagonal materials}

\author{J. Zhang}
\affiliation{Max-Planck-Institut f\"ur Festk\"orperforschung, Heisenbergstrasse 1, D-70569 Stuttgart, Germany} 
\affiliation{Quantum Matter Institute, University of British Columbia, Vancouver BC, Canada V6T 1Z4}

\author{Y.-H. Chan}
\affiliation{Institute of Atomic and Molecular Sciences, Academia Sinica, Taipei 10617, Taiwan}

\author{C.-K. Chiu}
\affiliation{Condensed Matter Theory Center and Joint Quantum Institute and Maryland Q Station, Department of Physics, University of Maryland, College Park, MD 20742, USA}

\author{M. G. Vergniory}
\affiliation{Donostia International Physics Center, P. Manuel de Lardizabal 4, San Sebasti‡n, 20018 Basque Country, Spain} 
\affiliation{Department of Applied Physics II, Faculty of Science and Technology, University of the Basque Country UPV/EHU, Apdo. 644, 48080 Bilbao, Spain}
\affiliation{IKERBASQUE, Basque Foundation for Science, Maria Diaz de Haro 3, 48013 Bilbao, Spain}
  
\author{L. M. Schoop}
\affiliation{Department of Chemistry, Princeton University, Princeton, NJ 08544, USA}

\author{A. P. Schnyder}
\affiliation{Max-Planck-Institut f\"ur Festk\"orperforschung, Heisenbergstrasse 1, D-70569 Stuttgart, Germany}

\begin{abstract}
Topological semimetals exhibit band crossings near the Fermi energy, which are protected by the nontrivial topological character of the wave functions. In many cases, these topological band degeneracies give rise to exotic surface states and unusual magneto-transport properties.
In this paper, we present a complete classification of all possible nonsymmorphic 
band degeneracies in hexagonal materials with strong spin-orbit coupling. 
This includes (i) band crossings protected by conventional nonsymmorphic symmetries, whose partial translation is within the invariant space of the mirror/rotation symmetry; 
and (ii) band crossings protected by off-centered mirror/rotation symmetries, whose partial translation is orthogonal to the invariant space.
Our analysis is based on (i) the algebraic relations obeyed by the symmetry operators
and (ii) the compatibility relations between irreducible representations at different high-symmetry points of the Brillouin zone.
We identify a number of existing materials where these nonsymmorphic nodal lines are realized. Based on these example materials,
we examine the surface states that are associated with the topological band crossings.  Implications for experiments and device applications
are briefly discussed.
\end{abstract}

\date{\rm\today}

\maketitle


\section{Introduction}

The study of semimetals with protected band degeneracies near the Fermi level has attracted a lot of interest in recent years,
due their unique transport and topological properties~\cite{chiu_RMP_16,volovikLectNotes13,armitage_mele_vishwanath_review,burkov_review_Weyl,yang_ali_review_ndoal_line}.
A multitude of new kinds of semimetals has been discovered, which includes Weyl semimetals~\cite{WanVishwanathSavrasovPRB11,BurkovBalentsPRB11,weyl_TaAs_ding_PRX_15}, Dirac semimetals~\cite{young_kane_rappe_Dirac_3D_PRL_12,dirac_sm_Na3Bi_fang_zhong_PRB_12,Dirac_SM_Cd3As2_fang_zhong_PRB_13},
and various types of nodal-line semimetals~\cite{xie_schoop_Ca3P2_apl_15,nodal_line_Yang,nodal_line_Yamakage,heikkila_volovik_JETP_11,schoop_zrsis}.
With these discoveries, it has become clear that
topology in combination with symmetry offers a useful organizing principle of semimetals.
Major strides have been made in developing topological classifications of semimetals using
 K-theory~\cite{horava_PRL_05,shiozaki_sato_classification_PRB_14,zhao_PT_PRL_16,zhao_PRL_13,chiu_schnyder_PRB_14}, symmetry-based indicators~\cite{pi_vishwanath_watanabe_nat_commun_17,khalaf_arXiv_17,song_fang_arXiv_17,fang_zhang_arXiv_17}, and compatibility relations between 
 irreducible representations~\cite{bradlyn_bernevig_nature_17,michel_zak_phys_reports_01,bradley_irreps,vergniory_PRE_17}.
These efforts have helped to group the semimetals into categories sharing common characteristics, and moreover, 
have pointed the way towards the discovery of new topological materials~\cite{chen_vishwanath_nat_phys_17,Bradlynaaf5037}.

Yet, a unified all-encompassing classification of semimetals, which combines the symmetries of the ten-fold way~\cite{chiu_RMP_16} with crystalline space group symmetries, is still lacking.
Such a complete enumeration of all possible semimetallic phases of matter would be a major milestone in the theory of solids. 
The derivation of a full classification represents, however, a formidable challenge, for which it will be likely necessary to not only apply, but also expand 
sophisticated mathematical tools, such as twisted equivariant K-theory~\cite{Freed2013}.
In order to make progress towards this goal, it is useful to first focus on a subset of all possible crystalline symmetries and to study the 
topological band degeneracies on a case-by-case basis. This can give insight into the mechanisms protecting band degeneracies and will
allow to build intuition about the classification patterns.

\begin{figure}[b!]
\centering
\includegraphics[width = 0.95\linewidth]{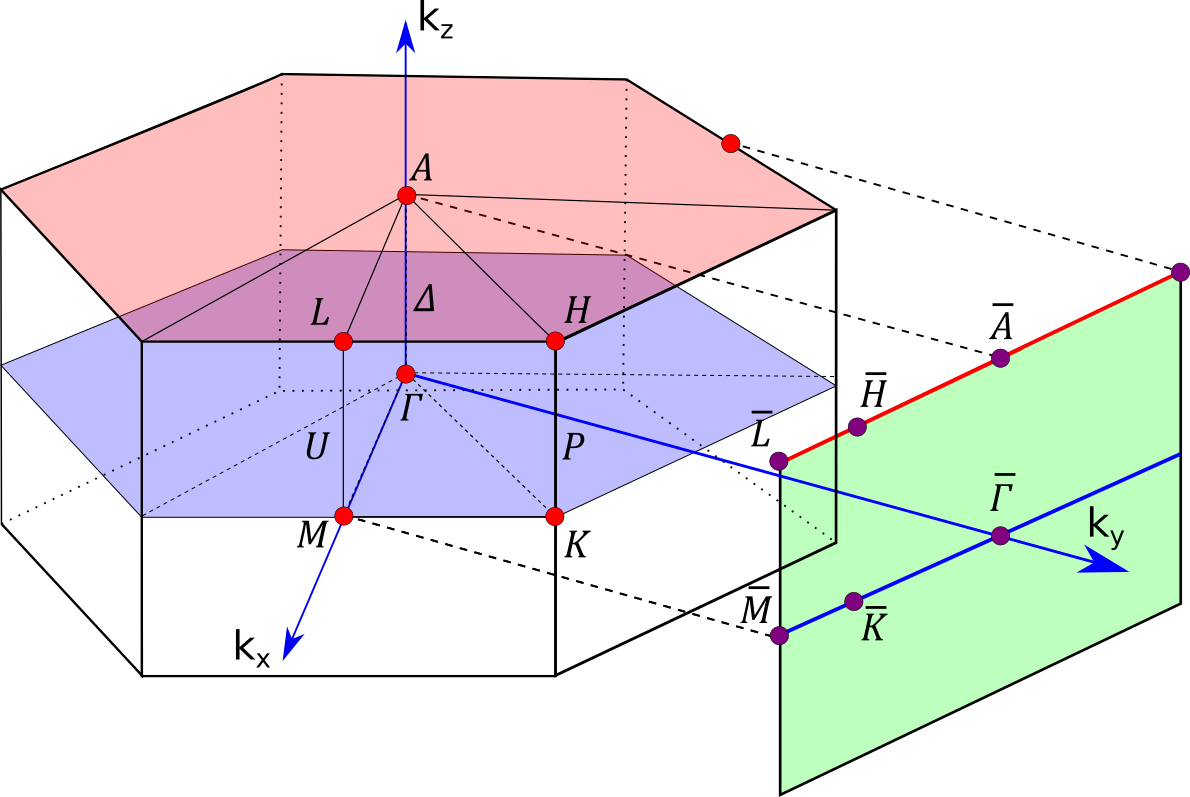}
\caption{ \label{mFig1}
Left: Bulk Brillouin zone for the hexagonal space groups, showing symmetry labels for high symmetry lines and points.
Right: Surface Brillouin zone for surfaces perpendicular to the (010) direction. 
}
\end{figure}

\begin{table*}[ht!]
\centering
\begin{ruledtabular}
\begin{tabular}{ l c c c c c c}  
\text{Space group} 
& 
Type-(i) nodal points
& 
Type-(i) nodal lines  
& 
Type-(ii) nodal lines   
& 
 Filling constraint & Materials \\ \hline
169 \ $(P6_1)$ & $\Gamma$$\Delta$A (12), MUL (4) &- & - & 12$\mathbb{N}$ & $\gamma$-In$_2$Se$_3$  \\ \hline
170 \ $(P6_5)$ & $\Gamma$$\Delta$A (12), MUL (4)  &-& - & 12$\mathbb{N}$ &-  \\ \hline
171 \ $(P6_2)$ & $\Gamma$$\Delta$A  (6)  &-& - &  6$\mathbb{N}$ &-\\ \hline
172 \ $(P6_4)$ & $\Gamma$$\Delta$A (6) &-& - & 6$\mathbb{N}$ &-\\ \hline
173 \ $(P6_3)$ & $\Gamma$$\Delta$A (4), MUL (4) &-& - & 4$\mathbb{N}$ &- \\ \hline
176  \ $(P6_3/m)$  &- & - &  $k_z = \pi$  &  4$\mathbb{N}$ & LaBr$_3$ \\ \hline
178 \ $(P6_1 22)$ & $\Gamma$$\Delta$A (12), MUL (4) &- & - & 12$\mathbb{N}$ & AuF$_3$ \\ \hline
179 \ $(P6_5 22)$ & $\Gamma$$\Delta$A (12), MUL (4) & - & -  & 12$\mathbb{N}$ & -\\ \hline
180 \ $(P6_2 22)$ & $\Gamma$$\Delta$A  (6) & - & -  & 6$\mathbb{N}$ & -\\ \hline
181 \ $(P6_4 22)$ & $\Gamma$$\Delta$A (6)  & - & - & 6$\mathbb{N}$ & -\\ \hline
182 \ $(P6_3 22)$ & $\Gamma$$\Delta$A (4), MUL (4) & - & - & 4$\mathbb{N}$ & -\\ \hline
188 \ $(P\bar{6}c2)$ & -  & $k_x k_z$-plane  & - &4$\mathbb{N}$ & LiScI$_3$ \\ \hline  
190 \ $(P\bar{6}2c)$ & - &    $k_x=\pi$ & - & 4$\mathbb{N}$ & ZrIrSn \\  \hline
\end{tabular}
\end{ruledtabular}
\caption{
Classification of nonsymmorphic band crossings in hexagonal materials with time-reversal symmetry and strong spin-orbit coupling.
The first column lists the nonsymmorphic space groups which exhibit topological band crossings.
The second and third columns indicate the high-symmetry lines and planes in which  type-(i) nodal points and nodal lines appear, respectively. 
For a definition of the coordinate system and high-symmetry labels in the hexagonal Brillouin zone see Fig.~\ref{mFig1}.
The values in the brackets denote the number of bands that form a connected group  with protected nodal points.
The high-symmetry planes which host type-(ii) nodal lines, protected by off-centered symmetries, are given in the fourth column.
The fifth column lists the electron fillings for which a band insulator is possible, see Sec.~\ref{sec_filling}. Here, $m \mathbb{N}$ denotes
the set $\{ m, 2m, 3m, \cdots \}$.
Some example materials which realize the predicted band crossings are listed in the last column. 
\label{mTab1}
}
\end{table*}

Motivated by these considerations, in this paper we classify topological band crossings in hexagonal materials with time-reversal symmetry and strong spin-orbit coupling. 
We focus on the role played by nonsymmorphic symmetries, which lead to symmetry-required band degeneracies along high-symmetry lines, 
or within high-symmetry planes, of the Brillouin zone (BZ)~\cite{michel_zak_PRB_99,zhao_schnyder_PRB_16,young_kane_non_symmorphic_PRL_15,alexandradinata_hourglass_surface_PRX_16,bzdusek_soluyanov_nature_16,furusaki_non_symmorphic_17,ryo_murakami_PRB_17,yang_furusaki_PRB_2017,fang_kee_fu_off_center_PRB_15,manes_PRB_12,Chiu_to_be_published}. 
While such an analysis has recently been performed to some degree
for cubic, tetragonal, and orthorhombic space groups (SGs)~\cite{bradlyn_bernevig_nature_17,bzdusek_soluyanov_nature_16,furusaki_non_symmorphic_17,ryo_murakami_PRB_17,yang_furusaki_PRB_2017}, nonsymmorphic hexagonal symmetries have not been considered before. Nonsymmorphic hexagonal SGs are special, because they contain sixfold screw rotations, which, as we will see, lead to multiple band crossings
with an accordion-like dispersion (cf.~Fig.~\ref{mFig3}). 
Another advantage of hexagonal materials is that they are usually easy to cleave, which makes them amenable to surface sensitive probes, such as scanning tunneling microscopy or photoemission 
spectroscopy. Also, note that nearly all known topological insulator materials are hexagonal or rhombohedral~\cite{ando_review_TI_JPSJ_13}.  
We study two different types of band degeneracies: (i)~twofold degenerate band crossings guaranteed by conventional
nonsymmorphic symmetries~\cite{bradlyn_bernevig_nature_17,bzdusek_soluyanov_nature_16,furusaki_non_symmorphic_17,ryo_murakami_PRB_17},
whose partial translation is parallel to the invariant space of the point-group symmetry; and (ii) fourfold degeneracies protected by off-centered symmetries,
whose partial translation is perpendicular to the invariant space of the point-group symmetry~\cite{yang_furusaki_PRB_2017,fang_kee_fu_off_center_PRB_15,Chiu_to_be_published}.
These twofold and fourfold degenerate topological band crossings can give rise to a range of exotic phenomena, including arc and drumhead surface states, quantum anomalies~\cite{burkov_anomalies_PRL_18,burkov_quantum_anomalies_long_arXiv_18,rui_arXiv_17}, 
anomalous magnetoelectric responses~\cite{bzdusek_soluyanov_nature_16}, and transverse topological currents~\cite{rui_arXiv_17}.

We classify both types of band crossings on a case-by-case basis, by computing the algebraic relations obeyed by the symmetry operators. For the SGs with \mbox{type-(i)} band degeneracies, we
also determine the compatibility relations between irreducible symmetry representations. We find that the band connectivities which follow from these compatibility relations are consistent with 
our results based on the algebraic relations of the symmetry operators. 
It should be noted, however, that \mbox{type-(ii)} band crossings cannot be derived from the compatibility relations in a straightforward manner.
Band crossings protected by off-centered symmetries are therefore missing in classification schemes that build on the theory of irreducible \mbox{(co-)representations} of 
crystallographic SGs~\cite{bradlyn_bernevig_nature_17,michel_zak_phys_reports_01,bradley_irreps,vergniory_PRE_17}. 
The results of our classification program are summarized in Table~\ref{mTab1}. We find that
there are two hexagonal SGs with type-(i) degeneracies forming Weyl nodal lines,
ten hexagonal SGs with type-(i) degeneracies forming Weyl nodal points, and
one hexagonal SG  with type-(ii) Dirac nodal lines.
Hence, in total there are 13 hexagonal SGs with nonsymmorphic band crossings. 
We note that these band crossings are required by the symmetries alone, regardless of the details of the particular crystal structure and composition.
In other words, the nonsymmorphic nodal lines cannot be annihilated while preserving the SG symmetries.
This is in contrast to accidental nodal lines~\cite{xie_schoop_Ca3P2_apl_15,nodal_line_Yang,nodal_line_Yamakage}, which are only perturbatively stable and can therefore be annihilated by large symmetry-preserving perturbations.
 
A secondary but equally important goal of this paper is to identify materials that exhibit the predicted band crossings.
For this purpose we use the Inorganic Crystal Structure Database (ICSD) from FIZ Karslruhe~\cite{ICSD_link} to look for materials
with the SGs listed in Table~\ref{mTab1}. We identify a number of suitable compounds, for example,  AuF$_3$ and $\gamma$-In$_2$Se$_3$, which
exhibit Weyl band crossing points with   accordion-like dispersions, and LaBr$_3$, which shows fourfold degenerate band crossings
with a star-like shape.
The topological properties of these band degeneracies manifest themselves at the surface
in terms of arc and drumhead surface states.
 To exemplify this, we compute the surface states of AuF$_3$ and show that there
  exist a number of arc states that connect the projected Weyl points of the accordion-like dispersion.

The remainder of the paper is organized as follows. 
In Sec.~\ref{mSec2} we classify band crossings protected by conventional nonsymmorphic symmetries. 
We consider both nodal points (Sec.~\ref{mSec2A}) and nodal lines (Sec.~\ref{mSec2B}). 
The band connectivities that follow from the compatibility relations between different irreducible representations are discussed
in Sec.~\ref{mSec2C}.
Sec.~\ref{mSec3}  is devoted to the classification of band crossings protected by off-centered symmetries.
In Sec.~\ref{mSec4} we present a few example materials which realize the predicted band crossings near the Fermi energy
and discuss their surface states.
Conclusions and perspectives are given in Sec.~\ref{mSec5}. 
Some technical details are relegated to Appendix~\ref{appendix_A} and Appendix~\ref{appendix_C}.
Appendix~\ref{appendix_B} contains additional band structure calculations of the example materials.

\section{Band crossings protected by conventional nonsymmorphic symmetries}
\label{mSec2}

Nonsymmorphic space groups contain symmetry operators  $G = \left\{ g | {\bm t} \right\}$
that combine point group symmetries $g$ with translations ${\bf t}$ by a fraction of a Bravais lattice vector~\cite{bradley_irreps}.
In the absence of additional symmetries  whose point of reference is different from $g$, the fractional translation ${\bf t}$ can be assumed to satisfy $g {\bf t} = {\bf t}$. 
This is because any component of ${\bf t}$ that is not invariant under $g$ can be removed by a suitable choice of reference for $g$.
In this section we focus on these types of nonsymmorphic symmetries, which we call ``conventional". (The case where there
are additional symmetries whose points of reference are different from $g$
 will be discussed in Sec.~\ref{mSec3}.)
Applying such a nonsymmorphic symmetry  $n$ times  yields an element of the lattice translation group~\cite{bradley_irreps,dresselhaus_group_theory}, i.e., 
\begin{eqnarray} \label{eq_G_power_n}
 G^n =  \left\{ g^n | n {\bf t} \right\} =  - p \,  T_{\bf a} , \quad  p \in \{1,2, \ldots, n -1 \} ,
\end{eqnarray}
where $g$ is  an $n$-fold point group symmetry and $T_{\bf a}$ is the translation operator for the Bravais lattice vector ${\bf a}$.
The minus sign on the right hand side of Eq.~\eqref{eq_G_power_n} originates from $g^n$, which equals $-1$ for Bloch electrons with non-negligible spin-orbit coupling. 

In the band structure of  nonsymmorphic materials, the operators  $G= \left\{ g | {\bm t} \right\}$  
can lead to the protection of band degeneracies in the $g$-invariant space of the BZ, which satisfies $g {\bf k} = {\bf k}$.  
In these   symmetry invariant lines and planes of the BZ, the Bloch states $\left| \psi_m ( {\bf k} ) \right\rangle$ can be constructed in such a way that they are simultaneous
eigenfunctions of both $G$ and the Hamiltonian. From Eq.~\eqref{eq_G_power_n} it follows that the eigenvalues of $G$ are 
\begin{eqnarray} \label{eq_G_eigenvals}
G \left| \psi_m ( {\bf k} ) \right\rangle 
=
e^{i \pi ( 2m+1) / n} e^{- i p {\bf k} \cdot {\bf a} / n } \left| \psi_m ( {\bf k} ) \right\rangle  ,
\end{eqnarray}
where $m \in \{ 0, 1, \ldots, n-1 \}$.
Due to the momentum dependent phase factor $e^{- i p {\bf k} \cdot {\bf a} / n }$ in Eq.~\eqref{eq_G_eigenvals} the eigensectors of $G$ can  be interchanged, 
as ${\bf k}$ is moved across the $g$-invariant space of the BZ. 
As a consequence, provided there are no additional degeneracies due to other symmetries,
pairs of bands must cross at least once  
within the invariant space.
This is the basic mechanism that leads to the protection of type-(i) band degeneracies~\cite{ryo_murakami_PRB_17,alexandradinata_hourglass_surface_PRX_16,furusaki_non_symmorphic_17,bzdusek_soluyanov_nature_16,young_kane_non_symmorphic_PRL_15,zhao_schnyder_PRB_16}, 
which we are now going to discuss for the hexagonal space groups.
The relevant nonsymmorphic symmetries that need to be considered for this purpose are:
 sixfold screw rotations $C_{6,p}$ of the form
\begin{eqnarray}
C_{6,p}  \;  : \;
(x, y, z) \to (x-y,x,z + \tfrac{p}{6} ) ( \tfrac{\sqrt{3}}{2} \sigma_0 - \tfrac{i}{2} \sigma_z   ) , \quad
\end{eqnarray}
and twofold glide mirrors $M$ of the form
 \begin{eqnarray} \label{glide_mirror_M}
M   \;  : \;
(x, y, z) \to (-x , y ,z  + \tfrac{1}{2} ) i \sigma_x ,
 \end{eqnarray}
where the Pauli matrices $\sigma_i$ operate in spin space.  
(See Fig.~\ref{mFig1} for the definition of the coordinate system.)
The screw rotations $C_{6,p}$ protect nodal points within rotation invariant lines, while the glide mirrors $M$ guarantee
the stability of nodal lines within mirror invariant planes.

\subsection{Weyl nodal points}
\label{mSec2A}

Nodal points occur along the $\Gamma$--$\Delta$--A and M--U--L lines of the hexagonal BZ, which are left invariant by the screw rotations $C_{6,p}$ and $C^3_{6,p}$, respectively.
Let us discuss these two cases separately.
It should be noted that nodal points discussed in this section can in certain cases be part of Weyl \emph{nodal lines}. An example of this is presented in Sec.~\ref{sec_AuF3}.

\begin{figure}[t!]
\centering
\includegraphics[width = \linewidth]{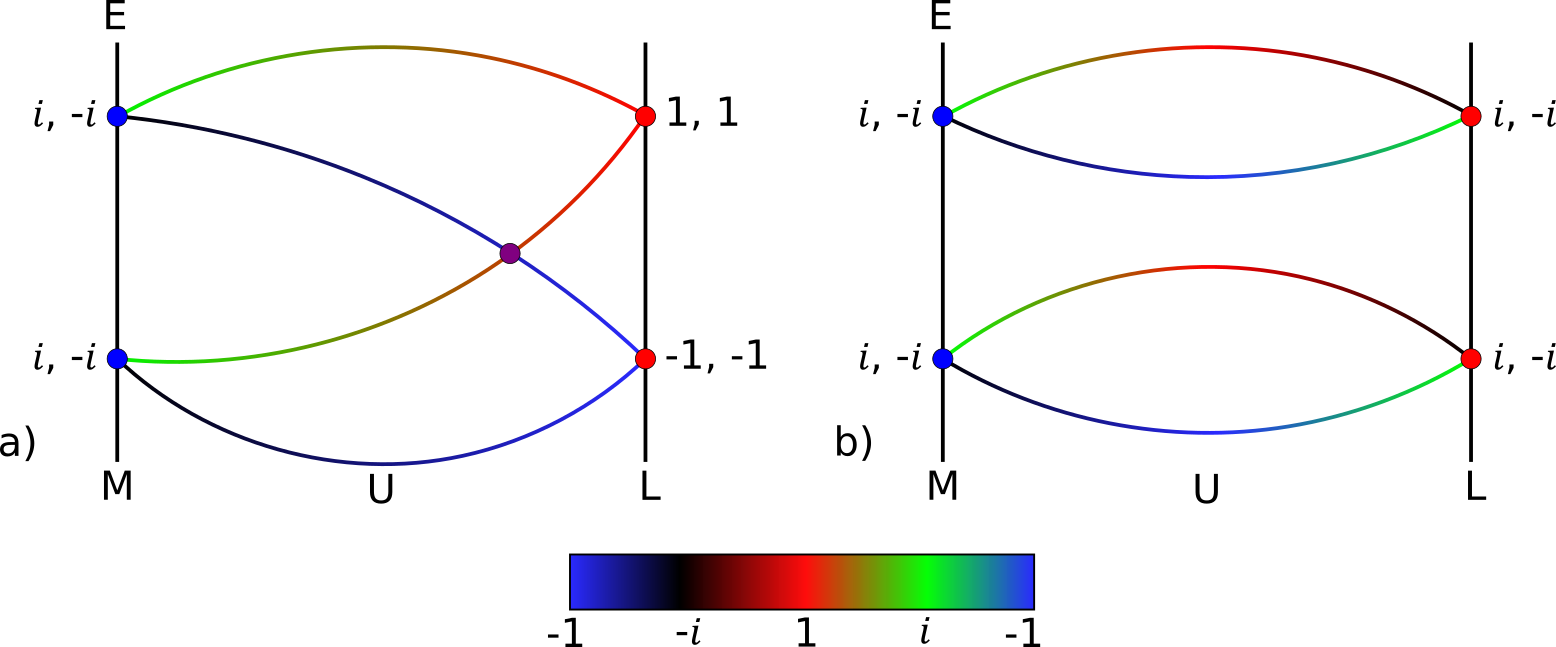}
\caption{\label{mFig2}
Band connectivity diagrams for the M--U--L line of the hexagonal BZ, which is left invariant
under the screw rotation $C^3_{6,p}$.
Panels (a) and (b) show the connectivities in the presence of a $C^3_{6,p}$ symmetry with $p$ odd and even, respectively.
The color scale indicates the screw rotation eigenvalues~\eqref{eq_C36p_ev} of the Bloch bands for (a) $p=1$ and (b) $p=2$.
The bands are Kramers degenerate at the time-reversal invariant momenta M and L.
}
\end{figure}

\begin{figure*}[t!]
\centering
\includegraphics[width = 0.1\linewidth]{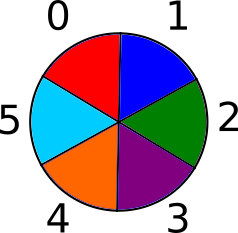}
\subfigure[]
{
	\includegraphics[width = 0.27\linewidth]{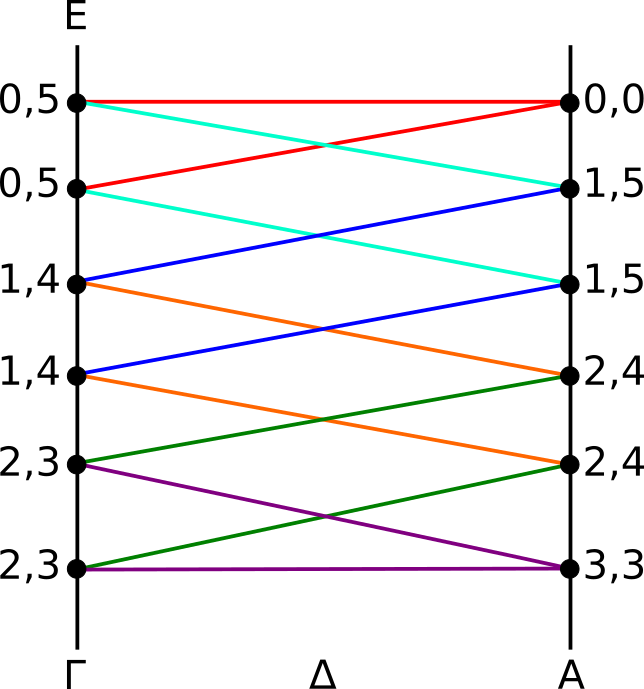}
}
\subfigure[]
{
	\includegraphics[width = 0.27\linewidth]{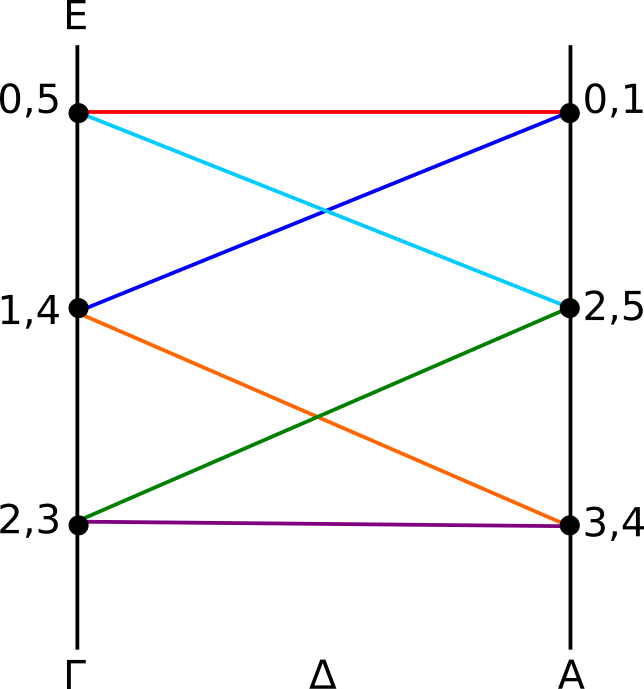}
}
\subfigure[]
{
	\includegraphics[width = 0.27\linewidth]{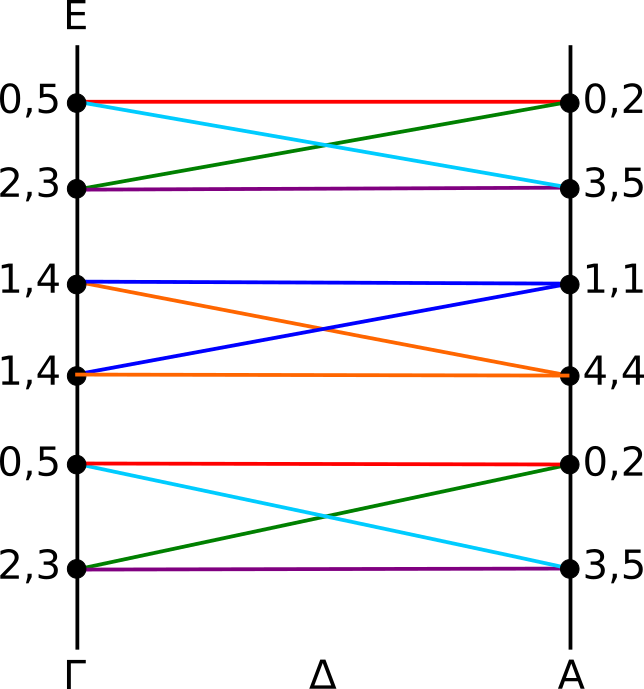}
}
\caption{ \label{mFig3}
Band connectivity diagrams along the $\Gamma$--$\Delta$--A line, which is left invariant under the sixfold screw rotation $C_{6,p}$.
Panels~(a), (b), and (c) show the connectivities in the presence of $C_{6,p}$ with (a) $p = 1$, (b) $p = 2$, and (c) $p = 3$. 
The  colors represent the $C_{6,p}$ eigenvalue labels $m$ of the Bloch bands $\left| \psi_m ( {\bf k} ) \right\rangle$, see Eq.~\eqref{eq_C6p_ev}. 
At the time-reversal invariant momenta $\Gamma$ and $A$ the bands form Kramers pairs (black dots).
The numbers next to the black dots indicate the $C_{6,p}$ eigenvalue labels $m$ of the two Kramers partners, cf.~Table~\ref{mTab2}.
For $p=4$ and $p=5$ the band connectivity is the same as for $p=1$ and $p=2$, respectively, although the band pairings at the time-reversal invariant momenta are different, cf.~Table~\ref{mTab2}.
}
\label{fig:GDAbands}
\end{figure*}

\subsubsection{\textrm{M--U--L}} \label{sec_MUL}
The M--U--L line is defined as the segment $\vk=(\pi, 0, k_z)$, with $k_z \in [0,\pi]$, 
which connects the two time-reversal invariant momenta (TRIMs) M and L along the $k_z$ direction, see Fig.~\ref{mFig1}. 
This line segment is left invariant under the twofold rotation $C^3_{6,p}$, up to a reciprocal lattice translation.
Hence, the Bloch bands $\left| \psi_m ( {\bf k} ) \right\rangle$ within this segment can be chosen to be simultaneous eigenstates of $C^3_{6,p}$
with eigenvalues
\begin{eqnarray} \label{eq_C36p_ev}
C^3_{6,p}  \left| \psi_\pm ( {\bf k} ) \right\rangle  = \pm i e^{-ipk_z/2}  \left| \psi_\pm ( {\bf k} ) \right\rangle ,
\end{eqnarray}
which follows from Eq.~\eqref{eq_G_eigenvals} with $n=2$ and ${\bf a} = (0,0,1)$.
Due to the presence of spin-orbit coupling, the energy bands $\left| \psi_m ( {\bf k} ) \right\rangle$ are in general non-degenerate,
except at TRIMs (i.e.,  at M and L), where time-reversal symmetry enforces Kramers degeneracies. 
From Eq.~\eqref{eq_C36p_ev}, we find that at  the M point $C^3_{6,p}$ has eigenvalues $\pm i$ for all $p$,
while at the L point the eigenvalues are $\pm i$ for $p$ even and  $\pm 1$ for $p$ odd.
At the M and L points time-reversal symmetry pairs up bands
whose $C^3_{6,p}$  eigenvalues are complex conjugate pairs.
Therefore,  Kramers partners at the M point have opposite $C^3_{6,p}$ eigenvalues, while at the L point they have the same eigenvalues for $p$ odd
and opposite eigenvalues for $p$ even. In the absence of additional symmetries, this leads
to the band connectivity diagrams shown in Fig.~\ref{mFig2}. 
For $p$ odd, we see that there are four bands forming a connected group, which must cross at least once, leading to Weyl point degeneracies with an hourglass dispersion~\cite{alexandradinata_hourglass_surface_PRX_16}.
For $p$ even, on the other hand, only two bands form a connected group without any symmetry-enforced crossings.

From the above considerations, we expect that materials in SGs containing $\Csix$ screw rotations with $p$ odd have
band crossings along the M--U--L line. However,   some of these SGs contain additional symmetries that lead to extra band degeneracies. 
In particular, mirror and inversion symmetries enforce extra  degeneracies at the L point, between  bands with $C^3_{6,p}$  eigenvalues $+1$ and $-1$.
Therefore, in the presence of   additional mirror or inversion symmetries the band crossing along the M--U--L line is no longer symmetry-required.
In conclusion, we find that only those materials with SGs containing a  $\Csix$ screw rotation with $p$ odd, but no inversion or mirror symmetry,
exhibit the Weyl degeneracies within the  M--U--L line.
These are the SGs with Nos.~169, 170, 173, 178, 179, and 182, as indicated in the second column of Table~\ref{mTab1}.

\subsubsection{$\Gamma$--$\Delta$--A} \label{sec_gamma_delta_A}

\begin{table}[b!]
\centering
\begin{ruledtabular}
\begin{tabular}{ c l l c }
&$p$ & \text{Pairings of $m$} \\ \hline
&0 & (0,5), (1,4), (2,3) \\ \hline
&1 & (0,0), (3,3), (1,5), (2,4) \\ \hline
&2 & (0,1), (2,5), (3,4) \\ \hline
&3 & (1,1), (4,4), (0,2), (3,5) \\ \hline
&4 & (0,3), (1,2), (4,5) \\ \hline
&5 & (2,2), (5,5), (0,4), (1,3) 
\end{tabular}
\end{ruledtabular}
\caption{ \label{mTab2}
This table specifies the way in which the Bloch bands $ \left| \psi_m ( {\bf k} ) \right\rangle$ pair up into Kramers partners at the A point of the BZ in materials with a sixfold screw rotation symmetry~$C_{6,p}$.
The pairing depends on $p$, i.e., the translation part of the screw rotation. 
The two $C_{6,p}$ eigenvalue labels $m$ of the Kramers pairs at the A point are indicated in the second column.
}
\end{table}

The path $\Gamma$--$\Delta$--A is defined as the segment  ${\bf k} = (0, 0, k_z)$, with $k_z \in [0, \pi ]$, which connects the $\Gamma$ point at the center of the BZ to the A point at the top surface of the BZ, see Fig.~\ref{mFig1}.
Since this path is invariant under the sixfold screw rotation $C_{6,p}$, we can label the Bloch states within $\Gamma$--$\Delta$--A by the eigenvalues of $C_{6,p}$, i.e.,
\begin{eqnarray}  \label{eq_C6p_ev}
C_{6,p}  \left| \psi_m ( {\bf k} ) \right\rangle =  e^{i \pi (2m+1)/6} e^{-i p k_z / 6} \left| \psi_m ( {\bf k} ) \right\rangle ,
\end{eqnarray}
where the eigenvalue label $m$ runs from 0 to 5, cf.~Eq.~\eqref{eq_G_eigenvals}.
Due to Kramers theorem,  time-reversal symmetry enforces degeneracies at the TRIMs $\Gamma$ and A, 
between bands whose $C_{6,p}$ eigenvalues are complex conjugate pairs. 
Using Eq.~\eqref{eq_C6p_ev}, we find that at the $\Gamma$ point bands with the eigenvalue labels 
(0, 5), (1, 4), and (2, 3) form Kramers partners, independent of $p$. At the A point, on the other hand,
the way in which the bands pair up into Kramers partners depends on the value of $p$, as specified in Table~\ref{mTab2}. 
We observe that for $p \ne 0$  the bands pair up in different ways at the $\Gamma$ and A points. As a consequence,
the Kramers pairs must switch partners between $\Gamma$ and A,
which, in the absence of additional symmetries, leads to a nontrivial band connectivity.
For $p=1$ and $p=5$ we find that there are twelve bands sticking together with an accordion-like dispersion, as shown in Fig.~\ref{mFig3}(a).
These twelve bands have to cross at least five times, forming five Weyl point degeneracies.
For $p=2$ and $p=4$ six bands stick together with a minimum number of two crossing points [Fig.~\ref{mFig3}(b)], while for $p=3$ four bands form at least one Weyl point degeneracy [Fig.~\ref{mFig3}(c)].
In general, one can show that for $2k$ bands forming a connected group, the minimum number of crossings is $k-1$. This is proven in Appendix~\ref{appendix_A} using mathematical induction. 
We note that the bands that cross at the Weyl points have different $C_{6,p}$ eigenvalues, which prevents hybridization between them.
In addition, the Weyl points are protected by a nonzero Chern number, which gives rise to Fermi arcs at the surface, as will  be demonstrated in Sec.~\ref{mSec4}.

\begin{figure}[t!]
\centering
\subfigure[]{
\includegraphics[width = 0.46\linewidth]{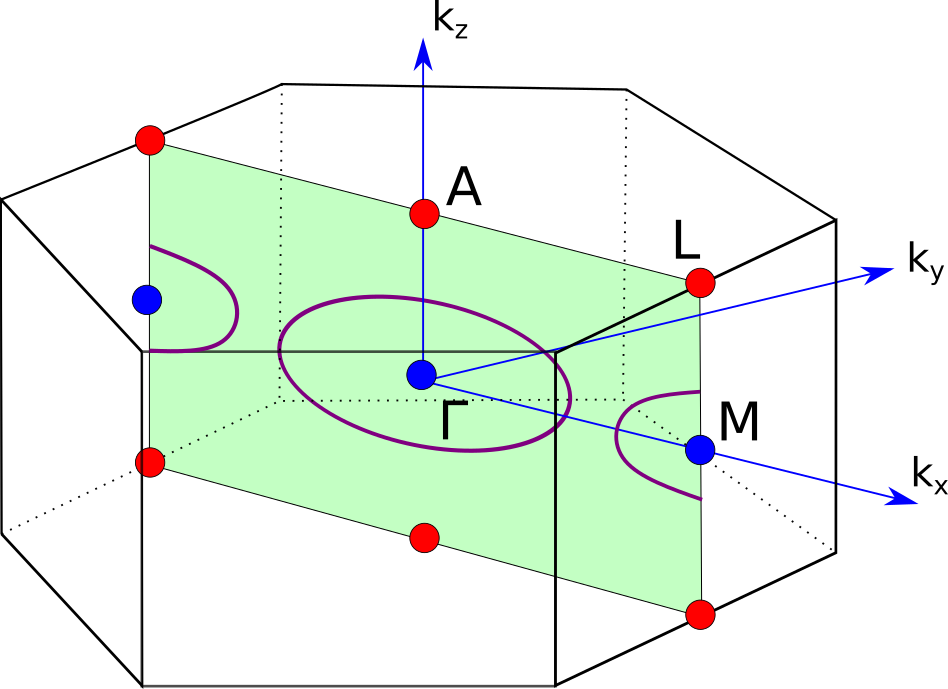}
}
\subfigure[]{
\includegraphics[width = 0.46\linewidth]{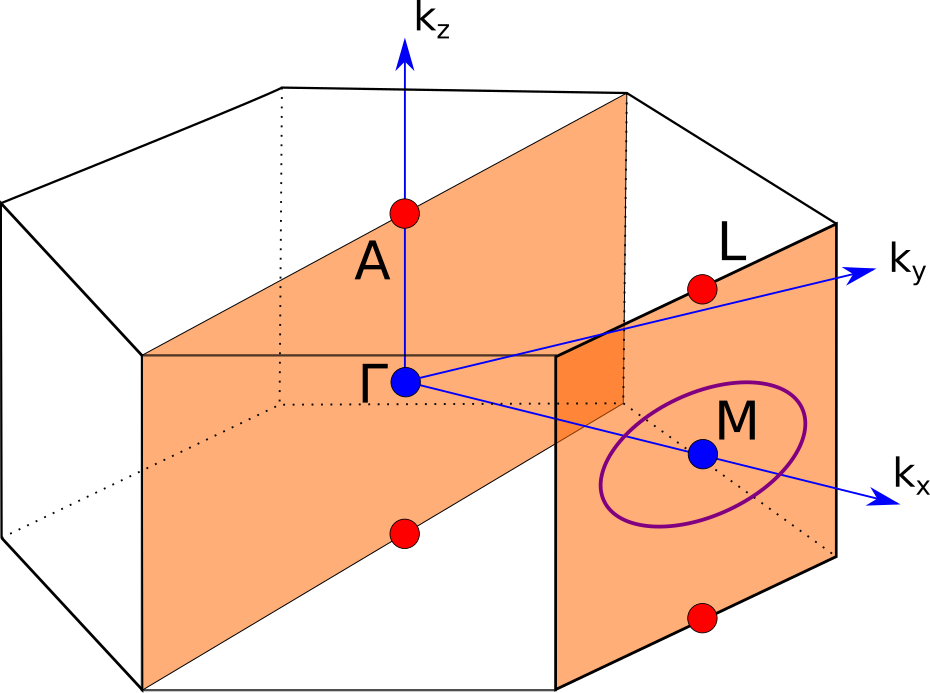}	
}
\caption{\label{mFig4}
Weyl nodal lines protected by glide mirror symmetries $M$~\eqref{glide_mirror_M} for (a) SG No.~188 and (b) SG No.~190.
The colored areas represent mirror-invariant planes. The bands are Kramers degenerate
at the time-reversal invariant momenta  (TRIMs) (blue and red dots).
In SG No.~188 the nodal lines enclose two TRIMs.
In SG No.~190 there is  at least one nodal line within the $k_x=\pi$ plane, enclosing a single TRIM.
There are different possibilities for the shape and connectivity of the nodal lines. Here we only show the
simplest one.
Note that  in SG No.~190
there is in general no symmetry-enforced nodal line within the $k_x = 0$ plane, since
the bands at the A point can be fourfold degenerate,
since the corresponding irreducible representations can have dimension $d=4$. 
}
\end{figure}
 
From the above analysis, it follows that materials in SGs with $C_{6,p}$ screw rotations can have protected Weyl point degeneracies along the $\Gamma$--$\Delta$--A line.
However, some of these SGs contain in addition to  $C_{6,p}$ also inversion or mirror symmetries, which create fourfould (or higher) degeneracies at the A point, thereby removing the
symmetry-enforced Weyl crossings. Therefore, only materials in SGs with $C_{6,p}$ symmetries, but no mirror or inversion symmetry, display the discussed
Weyl points at the $\Gamma$--$\Delta$--A line. The SGs that satisfy these criteria are Nos.~169-173 and 178-182, as listed in Table.~\ref{mTab1}.

\subsection{Weyl nodal lines}
\label{mSec2B}

Nodal lines protected by glide mirror symmetries $M$, Eq.~\eqref{glide_mirror_M},  occur within mirror-invariant planes of the hexagonal BZ~\cite{wang_hong_hourglass_PRB_17}.
Within these planes, the Bloch bands can be chosen to be eigenstates of the glide mirror operator $M$ with eigenvalues
\bee \label{mirror_EVs_two}
 M   \left| \psi_\pm ( {\bf k} ) \right\rangle =  \pm i e^{-ik_z/2}  \left| \psi_\pm ( {\bf k} ) \right\rangle,
\ee
which follows from Eq.~\eqref{eq_G_eigenvals}.
Mirror invariant planes contain either two or four TRIMs at which the Bloch bands are Kramers degenerate, see blue and red dots in Fig.~\ref{mFig4}.
It follows from Eq.~\eqref{mirror_EVs_two} that at the A and L points the $M$ eigenvalues are $+1$ and $-1$, while at the $\Gamma$ and M points they 
are $+ i$ and $-i$. 
At $\Gamma$ and M time-reversal symmetry enforces Kramers degeneracies between pairs with opposite 
$M$ eigenvalues (blue dots), while at  A and L  Kramers pairs are formed between bands with the same $M$ eigenvalues (red dots). 
This leads to a similar band connectivity as in Sec.~\ref{sec_MUL}, cf.~Fig.~\ref{mFig2}. That is, along any one-dimensional path within a mirror invariant plane, 
connecting $\Gamma$/M to A/L, there must be at least one band crossing.
Therefore, the mirror invariant planes must contain Weyl line degeneracies, as shown in Fig.~\ref{mFig4}.
The shape of these nodal line crossings is restricted by time-reversal symmetry,
which requires that each nodal line consists of pairs of Bloch states that are related by time-reversal symmetry. 
Due to these requirements each nodal line must enclose one or two TRIMs within the mirror plane.

The discussed nodal lines only occur if there are no extra degeneracies at the TRIMs due to additional symmetries. We find that the screw rotations $C_{6,p}$ or inversion lead
to extra degeneracies at the L and A points between bands with $M$ eigenvalues $+1$ and $-1$.
Hence, in the presence of these symmetries the L and M points
become fourfold degenerate, such that the line crossings are no longer enforced by symmetry. 
Therefore only materials in SGs with glide mirror symmetries, but no inversion or  $C_{6,p}$ symmetry, exhibit the predicted line crossings.
The only SGs that satisfy these criteria are Nos.~188 and 190~\footnote{For SG Nos.~184, 185, and 186 the bands with $M$ eigenvalues $+1$ and $-1$ at the A and L points are forced to be degenerate by extra mirror symmetries. Therefore, there is no symmetry-enforced crossing in SG Nos.~184, 185, and 186, even though they have a suitable glide plane as pointed out in Ref.~\cite{bzdusek_soluyanov_nature_16}.}. 
SG No.~188 ($P\bar{6}c2$) contains a glide mirror symmetry $M$ which leaves the $k_x k_z$-plane invariant, see Fig.~\ref{mFig4}(a). This plane contains four TRIMs: two where
the Kramers partners have the same $M$ eigenvalues (red dots); and two where they have opposite eigenvalues (blue dots). Using the above arguments, we find that there must 
exist nodal lines within the $k_x k_z$-plane,  
which enclose two of the four TRIMs, as shown in Fig.~\ref{mFig4}(a). 
The mirror glide symmetry $M$ in SG 190 ($P\bar{6}2c$), on the other hand, leaves two planes invariant, namely the $k_x=0$ and $k_x = \pi$  planes, see Fig.~\ref{mFig4}(b). 
Both of these mirror planes contain two TRIMs: one where the two Kramers partners
have the same $M$ eigenvalues (red dots); and one where they have opposite eigenvalues (blue dots). From our analysis it follows that both mirror planes should contain at least one nodal loop, which
either enclose the $\Gamma$/M or A/L points. 
However,  in SG No.~190 there exist four-dimensional irreducible representations at the A point, which makes the bands fourfold degenerate.
Thus, symmetry-enforced nodal loops generically occur only within the $k_x = \pi$ plane, but not within the $k_x=0$ plane.


\subsection{Compatibility relations between irreps}
\label{mSec2C}

 The symmetry enforced band crossings discussed in the previous two sections can also be derived from the compatibility relations
between irreducible representations (irreps)  
at different high-symmetry points (or lines) of the BZ~\cite{kruthoff_slager_PRX_17,elcoro_aroyo_JAC_17}. 
Here, we give a brief outline of this derivation focusing on SG 171 ($P6_5$), further details can be found in Appendix~\ref{appendix_C}.

The symmetries of electronic band structures with spin-orbit coupling and time-reversal symmetry are
described by double crystallographic SGs and their double-valued irreducible representations~\cite{bradley_irreps,elcoro_aroyo_JAC_17}. 
If we restrict the total band structure to a particular high-symmetry point ${\bf k}$ (or high-symmetry line) in the BZ, then the symmetries of the band structure are reduced
to a subgroup of the double SG, which is called the little group at ${\bf k}$. Since the Hamiltonian restricted to ${\bf k}$ commutes
with the corresponding little group, we can label its Bloch bands by the double valued irreps of the little group. 
Moving in a continuous way from a point with high symmetry (${\bf k}_1$, say) to a point with lower symmetry (${\bf k}_2$, say),
we find that the little-group irreps at these two points must be related to each other, as the little groups at ${\bf k}_1$ and ${\bf k}_2$ 
form a group-subgroup pair. In fact, a representation of the little group at ${\bf k}_2$ can be subduced from the little-group irreps at ${\bf k}_1$.
By decomposing this subduced representation into irreps, one obtains the compatibility relations between the irreps at ${\bf k}_1$ and ${\bf k_2}$~\cite{miller_lover_irreps,elcoro_aroyo_JAC_17}.
These compatibility relations then determine the connectivity of the Bloch bands in the BZ. 

\begin{table}[ht!]
\begin{subtable}
\centering
\begin{ruledtabular}
\begin{tabular}{ c | c | c | c | c | c | c}
Irrep\textbackslash Element & E & ${C}_{6,2}$ & ${C}_{6,2}^2$ & ${C}_{6,2}^3$ & ${C}_{6,2}^4$ & ${C}_{6,2}^5$ \\ \hline
$\overline{\Gamma}_7$ & 1 & -i & -1 & i & 1 & -i \\ \hline
$\overline{\Gamma}_8$ & 1 & i & -1 & -i & 1 & i \\ \hline
$\overline{\Gamma}_9$ & 1 & $e^{i5\pi/6}$ & $e^{-i\pi/3}$ & i & $e^{-i2\pi/3}$ & $e^{i\pi/6}$ \\ \hline
$\overline{\Gamma}_{10}$ & 1 & $e^{-i\pi/6}$ & $e^{-i\pi/3}$ & -i & $e^{-i2\pi/3}$ & $e^{-i5\pi/6}$ \\ \hline
$\overline{\Gamma}_{11}$ & 1 & $e^{i\pi/6}$ & $e^{i\pi/3}$ & i  & $e^{i2\pi/3}$ & $e^{i5\pi/6}$ \\ \hline
$\overline{\Gamma}_{12}$ & 1 & $e^{-i5\pi/6}$ & $e^{i\pi/3}$ & -i  & $e^{i2\pi/3}$ & $e^{-i\pi/6}$ \\ 
\end{tabular}
\end{ruledtabular}
\end{subtable}
\begin{subtable}
\centering
\begin{ruledtabular}
\begin{tabular}{ c | c | c | c | c | c | c}
Irrep\textbackslash Element & E & ${C}_{6,2}$ & ${C}_{6,2}^2$ & ${C}_{6,2}^3$ & ${C}_{6,2}^4$ & ${C}_{6,2}^5$ \\ \hline
$\overline{\textrm A}_{7}$ & 1 & $e^{-i\pi/6}$ & $e^{-i\pi/3}$ & -i & $e^{-i2\pi/3}$ & $e^{-i5\pi/6}$ \\ \hline
$\overline{\textrm A}_{8}$ & 1 & $e^{i5\pi/6}$ & $e^{-i\pi/3}$ & i & $e^{-i2\pi/3}$ & $e^{i\pi/6}$ \\ \hline
$\overline{\textrm A}_{9}$ & 1 & $e^{-i5\pi/6}$ & $e^{i\pi/3}$ & -i  & $e^{i2\pi/3}$ & $e^{-i\pi/6}$ \\ \hline
$\overline{\textrm A}_{10}$ & 1 & $e^{i\pi/6}$ & $e^{i\pi/3}$ & i  & $e^{i2\pi/3}$ & $e^{i5\pi/6}$ \\ \hline
$\overline{\textrm A}_{11}$ & 1 & i & -1 & -i & 1 & i \\ \hline
$\overline{\textrm A}_{12}$ & 1 & -i & -1 & i & 1 & -i \\ 
\end{tabular}
\end{ruledtabular}
\end{subtable}
\caption{ \label{irreps_G_A}
Double valued irreps of SG 171 ($P6_5$) without time-reversal symmetry
 at the $\Gamma$ and A points.
Here, we use the same convention as in Ref.~\onlinecite{elcoro_aroyo_JAC_17} for the labelling of the irreps.}
\end{table}

 We will now show how this works for the case of SG~171~($P6_5$).
For this SG the relevant high-symmetry points that we need to consider are the TRIMs  $\Gamma$ and A. These two TRIMs are connected
by the high-symmetry line $\Gamma$--$\Delta$--A,  which is left invariant under the screw rotation $C_{6,2}$ (cf.~Sec.~\ref{sec_gamma_delta_A}).
To determine the connectivity of the bands between the $\Gamma$ and A points, we   first derive the little-group irreps at $\Gamma$, A, and $\Gamma$--$\Delta$--A, and then
 study the compatibilities between them. Table~\ref{irreps_G_A} lists the calculated double-valued irreps at $\Gamma$ and A for SG~171 without time-reversal. 
 In order to construct time-reversal symmetric irreps we  pair complex conjugated irreps from Table~\ref{irreps_G_A} and form
 a direct sum out of them~\cite{bradley_irreps,elcoro_aroyo_JAC_17}.  In this way we obtain at the $\Gamma$ point the real irreps 
 $\overline{\Gamma}_7\overline{\Gamma}_8$, $\overline{\Gamma}_9\overline{\Gamma}_{12}$, and $\overline{\Gamma}_{10}\overline{\Gamma}_{11}$,
 while at the A point we get $\overline{\textrm A}_7\overline{\textrm A}_{10}$, $\overline{\textrm A}_8\overline{\textrm A}_9$, and $\overline{\textrm A}_{11}\overline{\textrm A}_{12}$.
Note that these real irreps are all two-dimensional, leading to a twofold degeneracy of the bands at $\Gamma$ and A, in agreement with Kramers theorem.
On the $\Gamma$--$\Delta$--A line, however, the symmetry is lower, as time-reversal is absent; only the screw rotation $C_{6,2}$ remains as a valid symmetry.
As a consequence, the little-group irreps at  $\Gamma$--$\Delta$--A are in general complex, as shown in Table~\ref{irreps_gamam_delta_A}.

\begin{table}[t!]
\centering
\begin{ruledtabular}
\begin{tabular}{ c | r | r}
Irrep\textbackslash Element & E & ${C}_{6,2}$\\ \hline
$\overline{\Delta}_7$ & 1 & $-i  e^{-ik_z/3}$ \\ \hline
$\overline{\Delta}_8$ & 1 & $i e^{-ik_z/3}$ \\ \hline
$\overline{\Delta}_9$ & 1 & $e^{i5\pi/6}e^{-ik_z/3} $\\ \hline
$\overline{\Delta}_{10}$ & 1 & $ e^{-i\pi/6}e^{-ik_z/3}$ \\ \hline
$\overline{\Delta}_{11}$ & 1 & $e^{i\pi/6}e^{-ik_z/3} $\\ \hline
$\overline{\Delta}_{12}$ & 1 & $ e^{-i5\pi/6} e^{-ik_z/3}$\\ 
\end{tabular}
\end{ruledtabular}
\caption{
\label{irreps_gamam_delta_A}
Double valued irreps of SG 171 ($P6_5$) at  the $\Gamma$--$\Delta$--A line.  
The irreps have momentum-dependent phases due to the partial translation of the ${C}_{6,2}$ symmetry. 
Note that the character $\chi$ of the symmetry elements ${C}^n_{6,2}$ can be inferred from the equation $\chi({C}^n_{6,2})$ = $\chi({C}_{6,2})^n$.
Here, we use the same convention as in Ref.~\onlinecite{elcoro_aroyo_JAC_17} for the labelling of the irreps.}
\end{table}

Next, we derive the compatibility relations between the little-group irreps at $\Gamma$, A and $\Gamma$--$\Delta$--A.
This can be achieved, in principle, by studying the subduction of the little-group irreps  at $\Gamma$ (or A) onto the little group
at $\Gamma$--$\Delta$--A~\cite{miller_lover_irreps}. Here, however, we can use a simpler method, which makes
use of the following relation between the characters $\chi$~\footnote{The character of a group irrep associates to each group element the trace of the corresponding irrep matrix.} 
 of the little-group irreps 
\bee \label{character_irrep_relation}
\chi [ \overline{D}_{l} \overline{D}_{\bar{l}} (g) ] = \sum_{i = 1}^{2} \chi [ \overline{\Delta}_{m_i} (g) ] ,
\ee
where $\chi [ \overline{D}_{l} \overline{D}_{\bar{l}} (g) ]$  is the character of the symmetry element $g$ for the real irrep $ \overline{D}_{l} \overline{D}_{\bar{l}}$
and $\{ \overline{\Delta}_{m_1}, \overline{\Delta}_{m_2} \}$ is the set of irreps that $ \overline{D}_{l} \overline{D}_{\bar{l}}$ decomposes into. 
Eq.~\eqref{character_irrep_relation} essentially follows from the fact that  the characters of each valid symmetry must be preserved, 
as we continuously move from   $\Gamma$ (or A)  to a point on the $\Gamma$--$\Delta$--A line.
Using Eq.~\eqref{character_irrep_relation} we find that the real irreps at $\Gamma$ must decompose into
\bee \label{compa_relation_Gamma}
 \overline{\Gamma}_7\overline{\Gamma}_8  &\rightarrow& \overline{\Delta}_7 + \overline{\Delta}_8 ,  \nonumber\\
 \overline{\Gamma}_9\overline{\Gamma}_{12} &\rightarrow& \overline{\Delta}_9 + \overline{\Delta}_{12} , \nonumber\\
 \overline{\Gamma}_{10}\overline{\Gamma}_{11} &\rightarrow& \overline{\Delta}_{10} + \overline{\Delta}_{11}  ,
\ee
while for the real irreps at A we have 
\bee  \label{compa_relation_A}
\overline{A}_7\overline{A}_{10} & \rightarrow & \overline{\Delta}_{8} + \overline{\Delta}_{11}, \nonumber\\
\overline{A}_8\overline{A}_9  & \rightarrow & \overline{\Delta}_{7} + \overline{\Delta}_{12},  \nonumber\\
\overline{A}_{11}\overline{A}_{12} & \rightarrow & \overline{\Delta}_{9} + \overline{\Delta}_{10}.
\ee 
These two sets of equations are the compatibility relations between the little-group irreps at
$\Gamma$, A, and $\Gamma$--$\Delta$--A. 

\begin{figure}[t!]
\centering
\includegraphics[width = 0.2\linewidth]{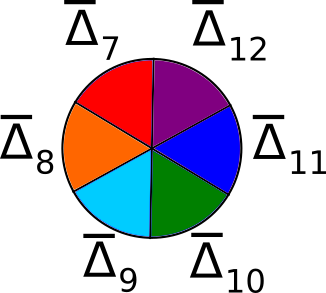}
\includegraphics[width = 0.6\linewidth]{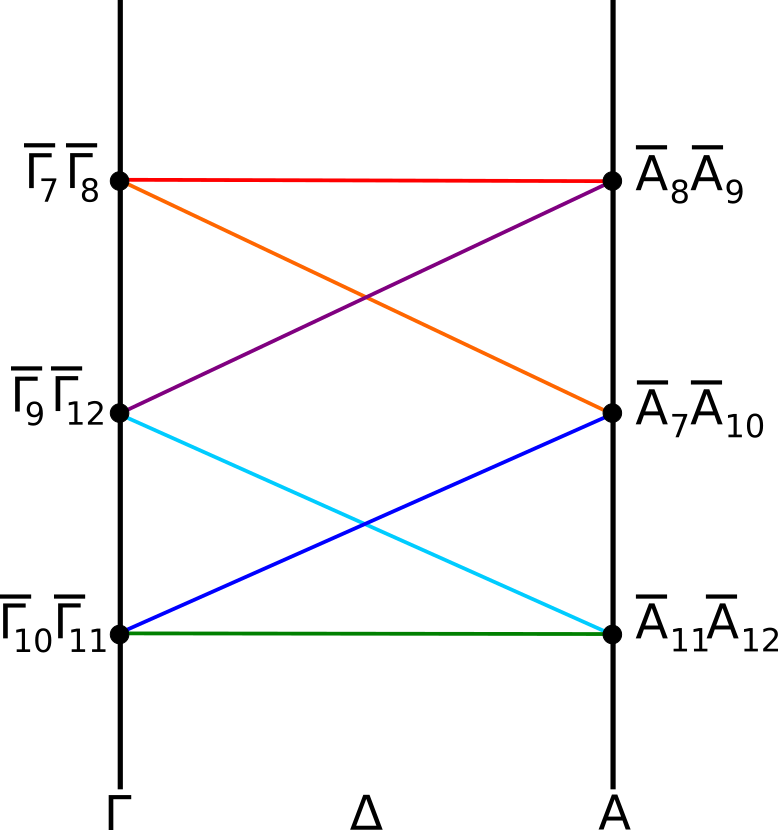}
\caption{
Band connectivity diagram for SG 171  ($P6_5$) along the $\Gamma$--$\Delta$--A line, which is left invariant under the screw rotation $C_{6,2}$.  
The bands at the TRIMs are labelled by real pairs of double valued irreps of $C_{6,2}$, using the convention of \cite{elcoro_aroyo_JAC_17}.
The colors indicate the double valued irreps at the $\Gamma$--$\Delta$--A line.
 }
\label{Graph171}
\end{figure}

The Bloch bands along $\Gamma$--$\Delta$--A must satisfy all of these compatibility relations. That is, as we move from $\Gamma$ to a point on  $\Gamma$--$\Delta$--A, Kramers pairs  must decompose
according to Eq.~\eqref{compa_relation_Gamma}, while, as we approach A, they must pair up according to Eq.~\eqref{compa_relation_A}. As a consequence, the little-group irreps of  
 $\Gamma$--$\Delta$--A  switch partners, as shown in Fig.~\ref{Graph171}. That is, the bands connect in a nontrivial way, with a minimum number of two crossings.
 This is in full agreement with Sec.~\ref{sec_gamma_delta_A}, cf.~Fig.~\ref{mFig3}(b). 

Using a similar approach as above, we have derived the compatibility relations for all  SGs of Table~\ref{mTab1} 
and constructed the corresponding band connectivity diagrams, see Appendix~\ref{appendix_C}.  We find that
the band connectivities obtained in this way fully agree with the derivation of Secs.~\ref{mSec2A} and~\ref{mSec2B}.
which is based on the algebraic relations obeyed by the symmetry operators.

\subsection{Filling constraints} \label{sec_filling}

It follows from band theory that, in the presence of time-reversal symmetry, a noninteracting band insulator can form only if the electron filling $\nu$ is an even integer, i.e., $\nu \in 2\mathbb{N}$~\footnote{The electron filling is defines as the number of electrons
per primitive unit cells.}. That is,  materials without strong correlations and $\nu \notin 2\mathbb{N}$ must necessarily be (semi-)metals.
However, in materials with nonsymmorphic symmetries these filling constraints for the existence of band insulators are tightened~\cite{michel_zak_phys_reports_01,young_kane_rappe_Dirac_3D_PRL_12}. 
I.e., nonsymmorphic symmetries forbid the existence of band insulators even when  $\nu \in 2\mathbb{N}$.
This is because nonsymmorphic symmetries generally enforce extra band crossings, leading to groups of more than two connected bands, as we have seen in Secs.~\ref{mSec2A} and~\ref{mSec2B}.
Thus, we can use the analysis of the above two sections to derive tight filling constraints for the corresponding SGs.
For example, for SG 169 we find that along the M--U--L line four bands form a connected group, while along the
$\Gamma$--$\Delta$--A line twelve bands form a connected group. Hence, for a material in SG 169 to be an insulator, these groups of bands must be fully filled,
i.e., the electron filling $\nu$ must be an element of  $ 4\mathbb{N}  \cap 12\mathbb{N}  =  12\mathbb{N}$.
Using similar arguments, tight filling constraints for all the SGs of Table~\ref{mTab1} can be derived,  see fifth column. 

Recently, the tight filling constraints for all 230 SGs with time-reversal symmetry were calculated by Watanabe \textit{et al.}~\cite{watanabe_vishwanath_PRL_16}, using  
compatibility relations between irreducible representations. Our analysis is in full agreement with Ref.~\cite{watanabe_vishwanath_PRL_16} and, moreover,
explicitly reveals both the nature and the topological protection of the enforced band crossings that lead to the tightened filling constraints.

\section{Band crossings protected by off-centered symmetries}
\label{mSec3}

Nonsymmorphic symmetries $G=\{ g | \bf{t}_\perp \}$ with a translation part ${\bf t}_\perp$ that is perpendicular to the $g$ invariant space
can be transformed into symmorphic symmetries by  a suitable choice of reference for $g$. For instance, a glide mirror symmetry $M = \{ m | {\bf t}_\perp \}$ with a translation
part ${\bf t}_{\perp}$ that is perpendicular to the mirror plane, can be transformed into a symmorphic mirror symmetry by shifting the origin by ${\bf t}_\perp / 2$. 
However, in the presence of a second symmetry  $G'=\{ g' | \bf{t}' \}$ with a reference point different from $G$, the two translation parts which are perpendicular to the point-group invariant
spaces cannot be removed both at the same time, since a shift of the origin affects  both $G$ and $G'$. 
A pair of two such symmetries are called ``off-centered" symmetries~\cite{yang_furusaki_PRB_2017,fang_kee_fu_off_center_PRB_15,Chiu_to_be_published}. 

We are now going to show how these off-centered symmetries in hexagonal SGs lead to the protection of fourfold degenerate nodal lines (i.e., Dirac lines), which we refer to as type-(ii) nodal lines
\footnote{In certain SGs off-centered symmetries can also protect type-(ii) nodal \emph{points}. This, however, is not the case for hexagonal materials with strong spin-orbit coupling.}.
The relevant off-centered symmetries that we need to consider for this purpose are: the glide mirror symmetry
$\widetilde{M}_z = \{ \widetilde{m}_z |  \frac{1}{2} \hat{z}  \}$
which transforms the spatial coordinates and the spin as
\begin{eqnarray} \label{def_glide_mriror_off_cent}   
\widetilde{M}_z \; : \; (x, y, z) \to (x,y,-z+1/2) i \sigma_z ,
\end{eqnarray}
together with the inversion symmetry $P$, which sends 
$  (x, y, z) \to (-x, -y, -z)$.
Since $\widetilde{M}_z^2 =  -1$, the eigenvalues of  $\widetilde{M}_z$ are $\pm i$.
Within the two $\widetilde{m}_z$ invariant planes, $k_x = 0$ and $k_x = \pi$, we can 
label the Bloch states $\left| \psi_m ( {\bf k} ) \right\rangle$, by the $\widetilde{M}_z$ eigenvalues, i.e., 
$\widetilde{M}_z \left| \psi_\pm ( {\bf k} ) \right\rangle =  \pm i \left| \psi_\pm ( {\bf k} ) \right\rangle$.
Applying the symmetry operators $\widetilde{M}_z$ and $P$ successively, we obtain the following commutation relation
\begin{eqnarray}
\widetilde{M}_z  P \left| \psi_\pm ( {\bf k} ) \right\rangle =e^{i k_z} P \widetilde{M}_z \left| \psi_\pm ( {\bf k} ) \right\rangle .
\end{eqnarray}
Thus, in the $k_z=0$ plane the two symmetry operators commute, while in the $k_z = \pi$ plane they anticommute. 
Now, since the time-reversal symmetry operator $T = i \sigma_y \mathcal{K}$ commutes with both $\widetilde{M}_z$ and $P$,
$\widetilde{M}_z$  anticommutes with $PT$ within the $k_z=\pi$ plane. Hence, the Kramers pair  $\left| \psi_\pm ( {\bf k} ) \right\rangle$
and $PT  \left| \psi_\pm ( {\bf k} ) \right\rangle$ have the same $\widetilde{M}_z$ eigenvalues for $k_z=\pi$, since
$\widetilde{M}_z \left[ P T  \left| \psi_\pm ( {\bf k} ) \right\rangle  \right] = -  P T \left[ \pm i  \left| \psi_\pm ( {\bf k} ) \right\rangle  \right] = \pm i  P T  \left| \psi_\pm ( {\bf k} ) \right\rangle$.
Therefore, if two bands with opposite $\widetilde{M}_z$ eigenvalues cross within the $k_z=\pi$ plane, they 
form a protected line crossing with fourfold degeneracy.

 \begin{figure}[t!]
\centering
\subfigure[]{
	\includegraphics[width = 0.48\linewidth]{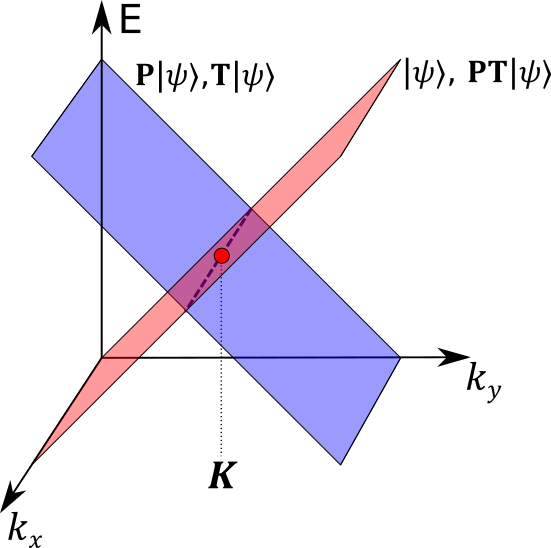}
}
\subfigure[]{
	\includegraphics[width = 0.45\linewidth]{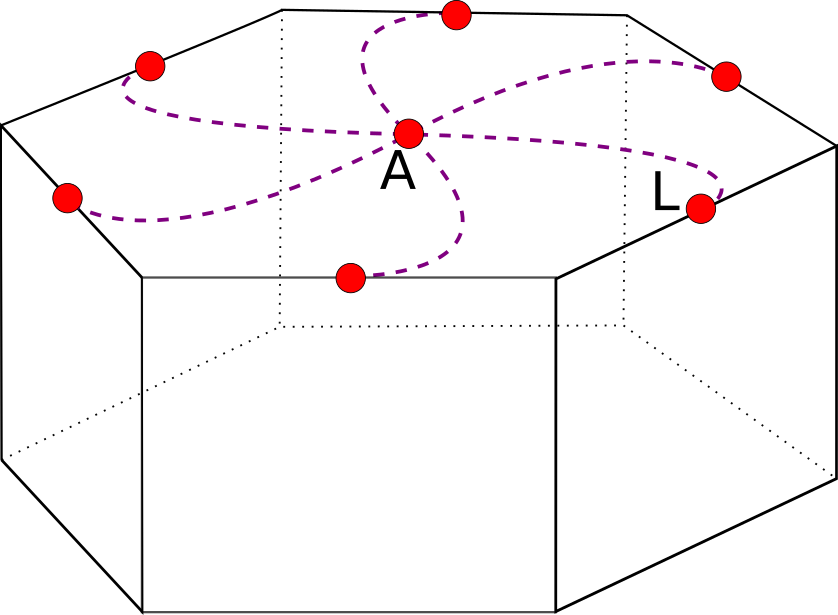}
}
\caption{ \label{mFig6}
(a) 
In materials with the off-centered symmetries $\widetilde{M}_z$ and $P$ [see Eq.~\eqref{def_glide_mriror_off_cent}], two Kramers degenerate bands with opposite $\widetilde{M}_z$ eigenvalues 
cross each other within the $k_z=\pi$ plane, forming a fourfold degenerate nodal line. The red and blue colors represent the $\widetilde{M}_z$ eigenvalues
$+i$ and $-i$, respectively. 
(b)
Fourfold degenerate Dirac nodal line protected by the off-centered symmetries,  connecting the A and L points of the hexagonal BZ. 
\label{fig_off_center_schematics}
}
\end{figure}
 
Such a fourfold degenerate nodal line is in fact required to exist by symmetry alone, i.e., it occurs in any material with the off-centered symmetries $( \widetilde{M}_z , P)$.
To see this, let us consider the degeneracies at the two TRIMs A and L within the $k_z = \pi$ plane. At these TRIMs the Bloch states
form quartets of four degenerate states
with the $\widetilde{M}_z$ eigenvalues
\bee
\widetilde{M}_z  \left| \psi_\pm ( {\bf K} ) \right\rangle &=& \pm i   \left| \psi_\pm ( {\bf K} ) \right\rangle , \nonumber\\
\widetilde{M}_z P  \left| \psi_\pm ( {\bf K} ) \right\rangle &=& \mp i  P \left| \psi_\pm ( {\bf K} ) \right\rangle , \nonumber\\
\widetilde{M}_z T  \left| \psi_\pm ( {\bf K} ) \right\rangle &=& \mp i T  \left| \psi_\pm ( {\bf K} ) \right\rangle , \nonumber\\
\widetilde{M}_z PT  \left| \psi_\pm ( {\bf K} ) \right\rangle &=& \pm i PT  \left| \psi_\pm ( {\bf K} ) \right\rangle , 
\ee
where ${\bf K} \in \{ \textrm{A}, \textrm{L} \}$.
These four Bloch states are mutually orthogonal to each other, since they either have opposite $\widetilde{M}_z$ eigenvalues or are Kramers partners.
As we move away from the TRIMs, the Bloch bands become twofold degenerate in general. We find, however, that the two energetically degenerate states $\left| \psi_\pm ( {\bf K} + {\bf k}) \right\rangle$ and $\left| \psi_\pm ( {\bf K} - {\bf k} ) \right\rangle$, which are mapped onto each other by $P$ (or by $T$), have opposite $\widetilde{M}_z$ eigenvalues. This leads to a band structure, whose $\widetilde{M}_z$ eigenvalues are inverted  with respect to ${\bf K}$, as shown in Fig.~\ref{mFig6}(a). Since the Bloch bands
are smooth functions of ${\bf k}$, 
each quartet of Bloch states at ${\bf K}$ must therefore be part of a fourfold degenerate nodal line connecting two TRIMs, as illustrated in Fig.~\ref{mFig6}(b). 
Note that this fourfold degenerate nodal line must be symmetric under $T$ and $P$ and all other point-group symmetries of the SG, but is otherwise free to move 
within the $k_z = \pi$ plane.  For this reason, the \mbox{type-(ii)} nodal lines in hexagonal systems are typically shaped like a star [see Fig.~\ref{mFig6}(b)].

From the above analysis, we expect that hexagonal materials in SGs containing both inversion and the glide mirror symmetry $\widetilde{M}_z$~\eqref{def_glide_mriror_off_cent} have
protected type-(ii) nodal lines.   
The SGs that fulfill these criteria are Nos.~176, 193, and 194 (cf.~Talbe~\ref{mTab1}).
However, for SG Nos.~193 and 194 additional symmetries force the nodal lines to be pinned to the high-symmetry lines connecting A to L and A to H, respectively.
Hence, in these cases the fourfold degeneracy is simply a consequence of the four dimensionality of the corresponding irreducible space-group representation, 
and hence the argument based on off-centered symmetries is not needed.

\section{Example Materials}
\label{mSec4}

To find example materials, we look for compounds crystallizing in one of the 13 SGs of  Table~\ref{mTab1} with heavy elements, showing large spin-orbit coupling.
For that purpose we perform an extensive search of the Inorganic Crystal Structure Database (ICSD) from FIZ Karslruhe~\cite{ICSD_link,aflow_citation},
focusing on binary compounds and simple ternary compounds. This search yields two materials with nodal points and two materials
with nodal lines, which we are now going to discuss in detail. 

For the four example materials we perform  
DFT calculations with the 
Vienna \textit{ab initio}  simulation package (VASP)~\cite{Kresse1996,Kresse1996-2}
using the projector augmented wave (PAW) method~\cite{Bloechl1994,Kresse1999} and the 
 PBE  functional~\cite{PBE_PRL_96} for the exchange-correlation energy.
As input for the DFT  calculations the experimental crystal structure of Refs.~\cite{zemva_AuF3_JACS_91,LIKFORMAN198091_In2Se3,zumdick99,meyer_gerd_LaBr3_89} was used.
In the main text  
we present only those features of the band structures that clearly show the predicted band crossings, while the full band structures are shown 
in Appendix~\ref{appendix_B}.
For one of the four materials, we also compute the surface states using  a Wannier-based tight-binding model~\cite{mostofi_wannier_08}
and an iterative GreenÕs function method~\cite{sancho_rubio_surface_green_functioin}.

\subsection{Materials with nodal points}
 
\begin{figure}[t!]
\includegraphics[width = 1.0\columnwidth]{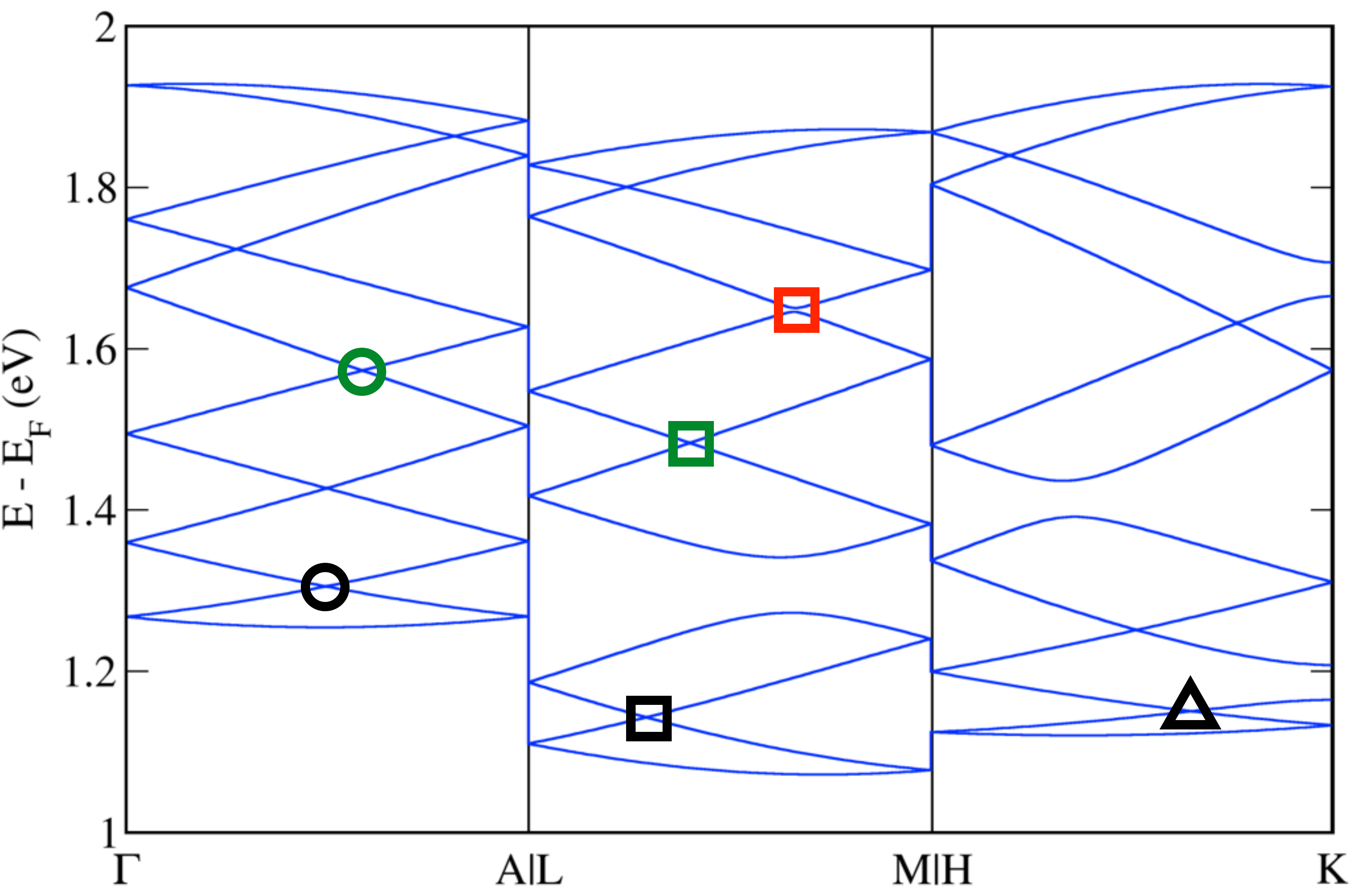}
\caption{
Electronic band structure of AuF$_3$ in SG $P6_122$ (No.~178). 
The band crossings along the  $\Gamma$--$\Delta$--A and M--U--L lines are symmetry-enforced
by the screw rotations $C_{6,1}$ and $C^3_{6,1}$, respectively (see Sec.~\ref{mSec2A}).
The red square in the middle panel indicates an avoided crossing. The black open 
symbols represent the Weyl points formed by the second and third lowest bands.
The green open symbols label the crossings of the sixth and seventh lowest bands, which are part of Weyl nodal lines. 
\label{fig_AuF3}
\label{mFig7}
}
\end{figure}

\paragraph{AuF$_3$---} \label{sec_AuF3}
Gold trifluoride AuF$_3$~\cite{zemva_AuF3_JACS_91} crystallizing in SG $P6_122$ (No.~178) is an example of a hexagonal material with Weyl point nodes along the  $\Gamma$--$\Delta$--A and M--U--L  lines (cf.~Sec.~\ref{mSec2A}). 
The first-principles band structure of AuF$_3$ displays a well separated group of twelve bands $\sim$1.5~eV above the Fermi energy $E_{\textrm{F}}$.
[Figs.~\ref{fig_AuF3} and~\ref{mFig_appendix_AuF3_In2Se3}(a)].
This group of bands shows the predicted Weyl points in a very clear way. 
Along the $\Gamma$--$\Delta$--A line, which is symmetric under the screw rotation $C_{6,1}$, we observe a group of twelve bands sticking together with an accordion-like dispersion, forming five crossings.
Along the  M--U--L  line, which is invariant under the screw rotation $C^3_{6,1}$, there are groups of four connected bands, 
with an hourglass dispersion and a single crossing point.
(Note that the two bands marked by the red square form an avoided crossing.)
This is in complete agreement with the theoretical band connectivity diagrams of Figs.~\ref{mFig3}(a) and~\ref{mFig2}(a), respectively. 
The occupied states below the Fermi level (see Fig.~\ref{mFig_appendix_AuF3_In2Se3}(a) in Appendix~\ref{appendix_B}) exhibit the same band connectivity as the unoccupied ones in Fig.~\ref{fig_AuF3}.
That is, along the $\Gamma$--$\Delta$--A and M--U--L  lines, there are
groups of 12$n$ and 4$n$ connected bands, respectively, which form a large number of Weyl points (and possibly also Weyl lines) protected by the screw rotation symmetries.

The topological stability of all these Weyl points is ensured by quantized Chern numbers, which endow the Weyl points with
definite chiralities $\nu$. Interestingly, the chiralities of the Weyl points of AuF$_3$ can be infered from symmetry alone, at least
up to some overall signs. This can be achieved using the following three observations.
First, the chiralities $\nu$ must be either $+1$ or $-1$, since all the bands cross linearly and not quadratically~\cite{huang_hasan_double_Weyl_PNAS_16}.
Second, the chiralities of all the Weyl points formed by one pair of bands must add up to zero, due to the fermion doubling theorem~\cite{NIELSEN198120}.
Third, Weyl points which are mapped onto each other under space group symmetries must have the same charilities.
Using these three observations, let us, as an example, analyze the  chiralities  of the Weyl points formed by the second and third lowest  bands
in Fig.~\ref{mFig7} (black open symbols).
How the bands connect across the entire BZ can be deduced from Fig.~\ref{mFig_appendix_AuF3_In2Se3} shown in Appendix~\ref{appendix_B}.
Looking at the hexagonal BZ in Fig.~\ref{mFig1}, we see that there are three Weyl points at the M--U--L lines with chirality $\nu_{\textrm{MUL}}$, two Weyl points at the K--P--H lines with chirality $\nu_{\textrm{KPH}}$,
and one Weyl point at the $\Gamma$--$\Delta$--A line with chirality $\nu_{\Gamma \Delta \textrm{A}}$.
Hence, the chiralities of these six Weyl points must obey the equation
$3 \nu_{\textrm{MUL}}  + 2 \nu_{\textrm{KPH}} + \nu_{\Gamma \Delta \textrm{A}}  = 0$, which, up to an overall sign, fully determines the chiralites, i.e., 
$( \nu_{\textrm{MUL}},    \nu_{\textrm{KPH}},   \nu_{\Gamma \Delta \textrm{A}} ) = ( +1, -1,-1)$
\footnote{It should be noted that the second and third lowest conduction bands can, in principle, also form accidental Weyl points, i.e., Weyls points aways from high-symmetry lines, whose existence is not enforced by symmetry.
These accidential Weyl points always come in opposite chirality pairs, such that their contribution to the equation for the chiralities cancels out.}
.
Similar arguments can be made for all the other Weyl points of AuF$_3$ and in fact for any material with
these type-(i) Weyl points (cf.~Table~\ref{mTab1}).
A peculiar case are the crossings formed by the sixth and seventh lowest bands in Fig.~\ref{mFig7} (green open symbols).
We find that these bands cross only on the  $\Gamma$--$\Delta$--A and M--U--L lines, but not along the K--P--H path. Therefore,
the chiralities must satisfy  $3 \nu_{\textrm{MUL}} + \nu_{\Gamma \Delta \textrm{A}}  = 0$, which paradoxically has no solution.
The resolution to this conundrum, is that the band crossings marked by the green symbols are not actually Weyl points, but rather Weyl nodal lines, whose chirality is illdefined.
These Weyl nodal lines connect the M--U--L path to the $\Gamma$--$\Delta$--A path in a sixfold star-shaped pattern, which is confirmed by our \emph{ab-initio} DFT calculations (not shown).

 \begin{figure}[t!]
\includegraphics[width = 1.0\columnwidth]{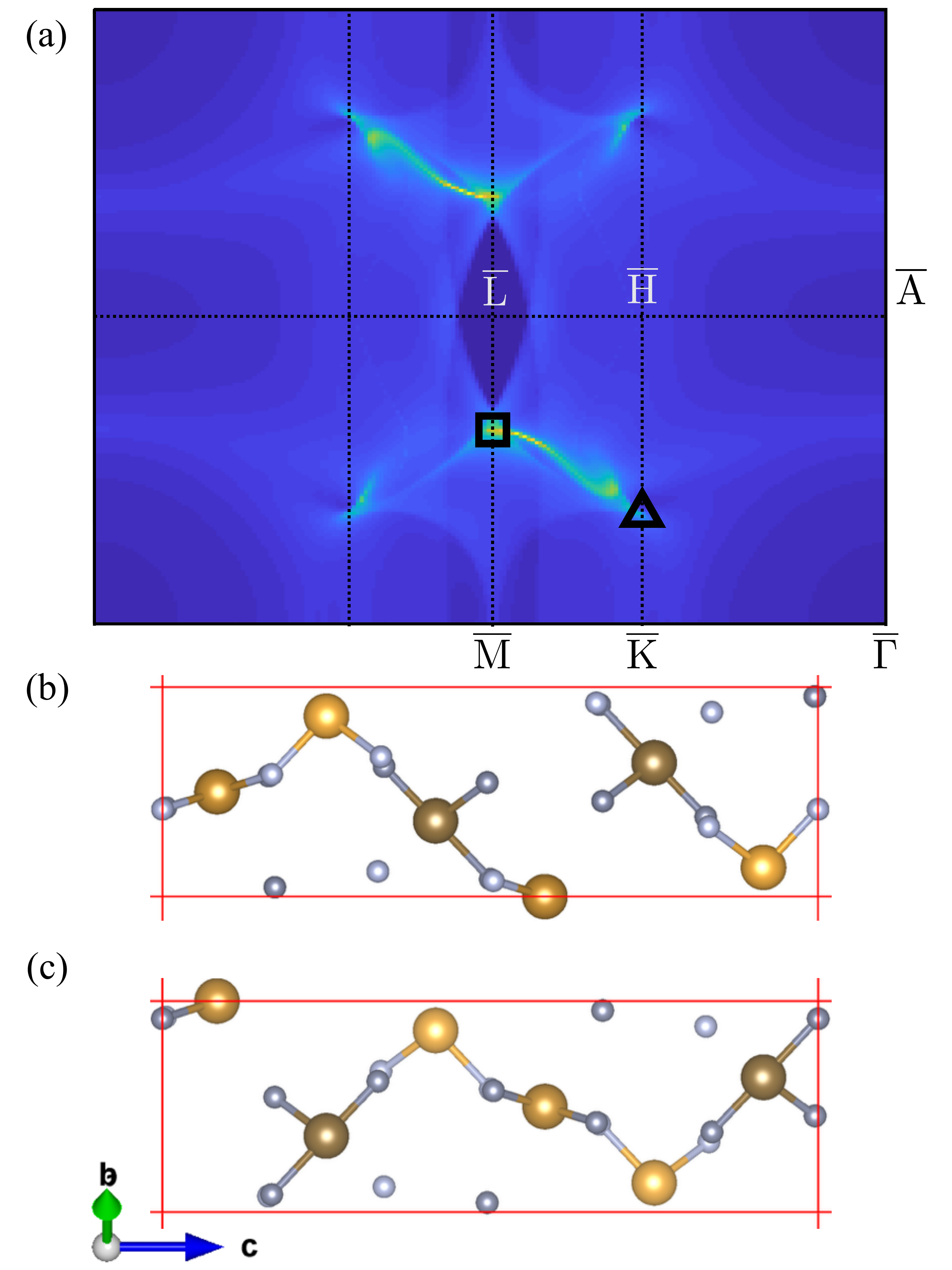}
\caption{ \label{fig_AuF3_surface}
(a) Momentum-resolved surface density of states for the (010) face of AuF$_3$ with bottom termination at the energy $E - E_{\textrm{F}} = 1.14$~eV.
Yellow and blue correspond to high and low density, respectively. 
The black square and triangle indicate the location of the Weyl points along the M--U--L and K--P--H lines, respectively (cf.~Fig.~\ref{mFig7}).
There are two arc states connecting the two Weyl points.
 Panels (b) and (c) show the Au   and F  atoms in the the two outermost layers of the top and bottom (010) terminations, respectively. 
The Au (F) atoms in the first and second layers are color by light and dark yellow (grey), respectively.
}
\end{figure}

 \begin{figure}
\includegraphics[width = 1.0\columnwidth]{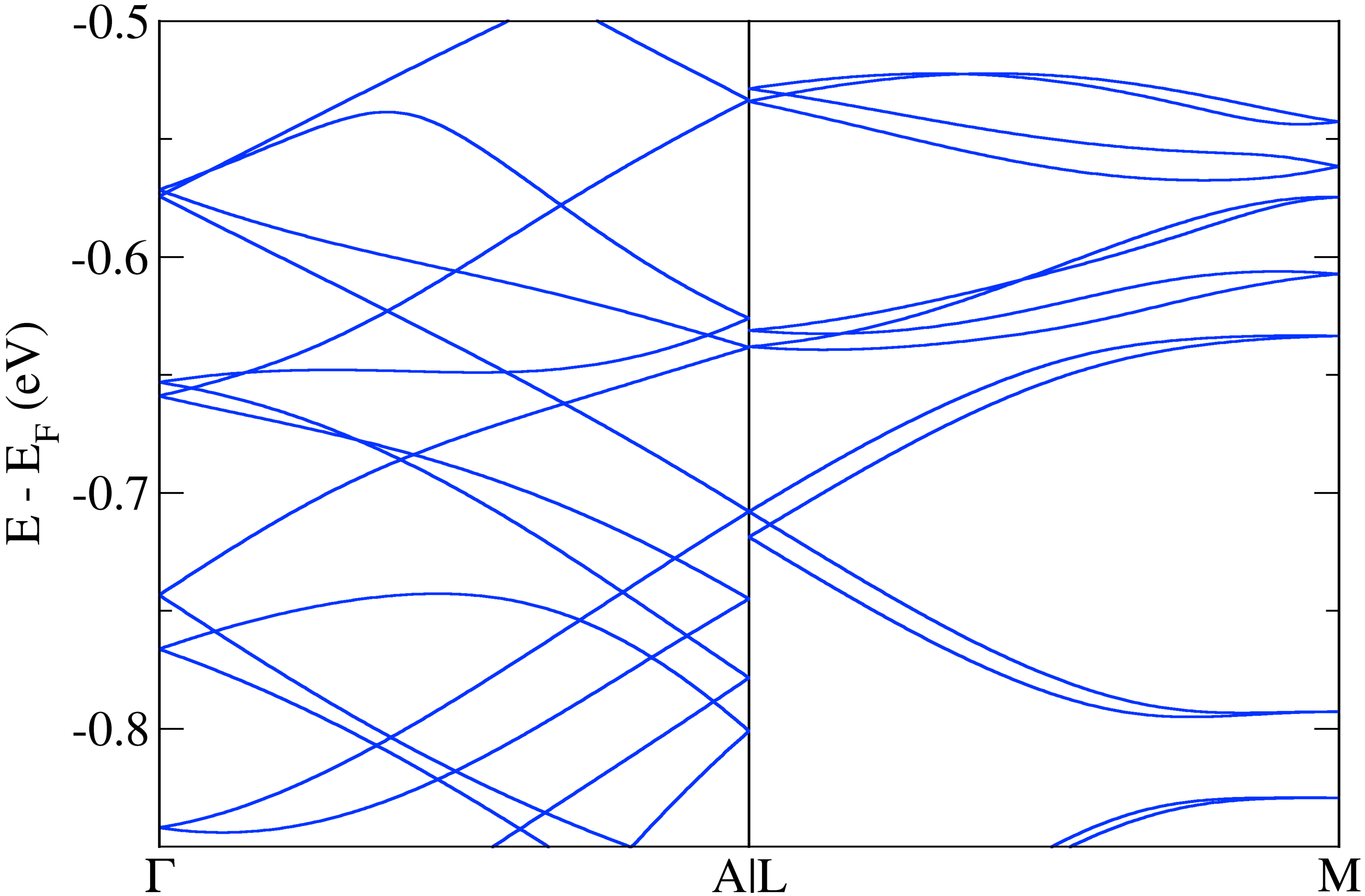} 
\caption{ \label{fig_In2Se3}
First-principles band structure of  $\gamma$-In$_2$Se$_3$ in SG $P6_1$ (No.~169). Weyl nodal points protected by the screw rotations
$C_{6,1}$ and $C^3_{6,1}$ occur along the $\Gamma$--$\Delta$--A and M--U--L lines, respectively (see Sec.~\ref{mSec2A}).
}
\end{figure}

The nontrivial topology exhibited by the Weyl points of AuF$_3$ manifests itself at the surface through arc states, which connect Weyl points with opposite chiralities. 
This is shown in Fig.~\ref{fig_AuF3_surface}, which displays
the surface density of states of a semi-infinite AuF$_3$ slab with (010) face at the energy $E - E_{\textrm{F}} = 1.14$~eV.
The corresponding surface BZ is shown on the right of Fig.~\ref{mFig1}.
The crystal structure of AuF$_3$ allows for two different (010) terminations, which are displayed
in Figs.~\ref{fig_AuF3_surface}(b) and~\ref{fig_AuF3_surface}(c). In Fig.~\ref{fig_AuF3_surface}(a) we show the surface density of states for the bottom termination; 
the corresponding plot for the top termination is presented in Fig.~\ref{top_surface_state_AuF3} of Appendix~\ref{appendix_B}.
We observe that there are two arc states connecting the projected Weyl points of the M--U--L and
K--P--H lines, which are marked by the black square and triangle, respectively.
Similar arc states occur for all the other Weyl points along the $\Gamma$--$\Delta$--A and M--U--L lines,
also in the occupied bands.

While the band connectivity and the associated Weyl points of AuF$_3$ are very interesting from a theoretical point of view, 
this material is rather difficult to handle experimentally, as it is highly reactive.
Moreover, AuF$_3$ is insulating with a large band gap of $\sim2$~eV, which makes it nearly impossible to probe the bulk Weyl points
and arc surface states using tunneling spectroscopy, photoemission, or transport experiments.
Therefore, in order to experimentally confirm the predicted band crossings, it is more feasible to
 instead consider the $\gamma$ phase of In$_2$Se$_3$.

\paragraph{In$_2$Se$_3$---}
In$_2$Se$_3$ exists in several polymorphic forms~\cite{JI20132517}. In one of them, the so called $\gamma$ phase, 
In$_2$Se$_3$ crystallizes in the space group $P6_1$ (169)~\cite{LIKFORMAN198091_In2Se3}. 
Hence, according to our classification Table~\ref{mTab1},
the band structure of $\gamma$-In$_2$Se$_3$ exhibits Weyl points along the 
$\Gamma$--$\Delta$--A and M--U--L lines, which are protected by  the screw rotations
$C_{6,1}$ and $C^3_{6,1}$, respectively. 
Fig.~\ref{fig_In2Se3} displays the \emph{ab initio} band structure of $\gamma$-In$_2$Se$_3$,
which clearly shows these Weyl points. We observe groups of four connected bands along  the M--U--L direction, which cross
at least once, similar to the band connectivity diagram~\ref{mFig2}(a).
Along the $\Gamma$--$\Delta$--A line there are groups of
12$n$ connected bands, which form at least $6n-1$ Weyl points [Fig.~\ref{fig_In2Se3} and~\ref{mFig_appendix_AuF3_In2Se3}(b)].
Similar to AuF$_3$, these Weyl points lead to arc surface states, due to the bulk boundary correspondence. 
Here, however, the arc states  
are much closer to the Fermi energy, which makes it possible to observe them using, e.g., angle-resolved photoemission or scanning tunneling spectroscopy.
Moreover, $\gamma$-In$_2$Se$_3$ can have p-type defects that would move the Fermi level closer to the Weyl points.
This would allow to measure the topological transport signatures of the Weyl points, e.g.,  anomalous (magneto-)transport 
properties due to the chiral anomaly~\cite{armitage_mele_vishwanath_review,burkov_review_Weyl}.
In addition, $\gamma$-In$_2$Se$_3$ is expected to exhibit large anomalous Hall effects,
since all the bands carry a non-zero Berry curvature, which is especially large close to the Weyl points.

\subsection{Materials with nodal lines}

Here, we present two materials with symmetry enforced nodal lines: ZrIrSn, which has twofold degenerate Weyl nodal lines [type-(i)] and
LaBr$_3$, which exhibits fourfold degenerate Dirac nodal lines [type-(ii)].

\begin{figure}
\includegraphics[width = 1.0\columnwidth]{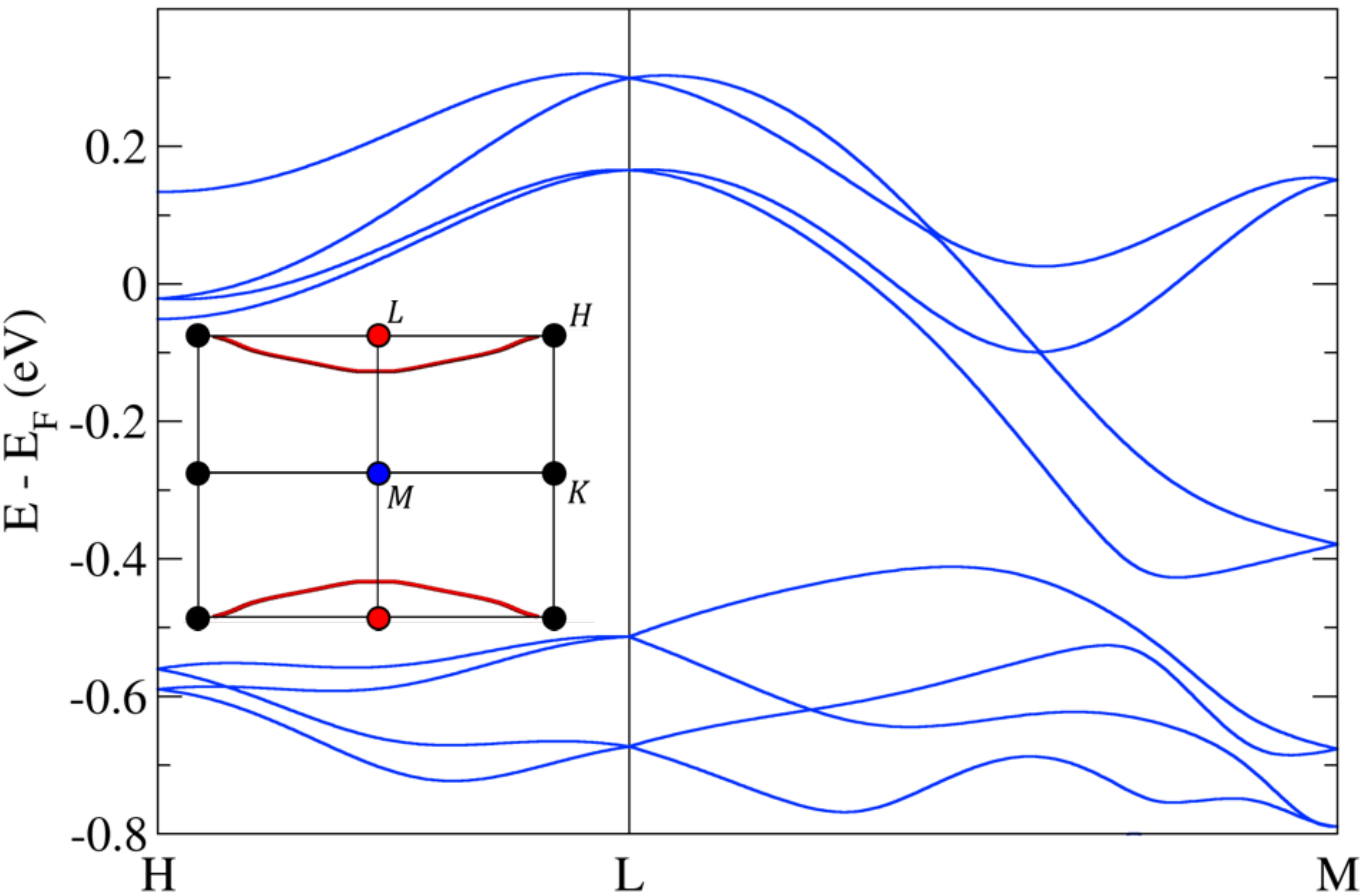}
\caption{ \label{fig_ZrIrSn}
Electronic band structure of ZrIrSn in SG $P\bar{6}2c$ (No.~190).
The band crossings along the M--U--L path are part of Weyl nodal lines
within the $k_x = \pi$ plane, which are protected by the glide mirror symmetry $M_x$.
These Weyl nodal lines enclose one of the two time-reversal invariant momenta M or L, as shown in 
the inset for the crossing near $E \simeq -0.64$~eV.
\label{fig_IrSnZr}}
\end{figure}

\paragraph{ZrIrSn---}
ZrIrSn crystallizing in SG $P\bar{6}2c$ (No.~190)~\cite{zumdick99} is an example of a hexagonal material with Weyl nodal lines protected by a  mirror glide symmetry.
In Fig.~\ref{fig_ZrIrSn} we present the first-principles band structure of ZrIrSn.
Along the M--U--L line, which is invariant under the mirror glide symmetry $M_x$, we observe groups of four connected bands, which cross each other at least once.
Since these crossings must occur for any path within the $k_x = \pi$ plane connecting M to L  (cf.~Sec.~\ref{mSec2B}), they form Weyl nodal lines.
The shape of the Weyl nodal line for the crossing near $E \simeq -0.64$~eV is shown in the inset of Fig.~\ref{fig_ZrIrSn}. All the other Weyl nodal lines have similar shapes
and enclose one of the TRIMs M or L in the $k_x = \pi$ plane.
We emphasize that all the bands within the $k_x = \pi$ plane form such Weyl nodal lines, since their existence is enforced by the mirror glide symmetry.
The topological properties of these Weyl nodal lines are characterized by a nonzero Berry phase~\cite{nodal_line_Yang},
which, by the bulk-boundary correspondence, leads to drumhead states at the surface of ZrIrSn.
Moreover, due to the absence of inversion, the bands in ZrIrSn carry a nonzero Berry curvature, 
which is particularly large close to the Weyl nodal lines.
 In slightly doped samples of ZrIrSn  this should give rise to anomalous transport properties, such as, e.g., large anomalous Hall effects or 
anomalous magnetoelectric responses~\cite{bzdusek_soluyanov_nature_16}.

\begin{figure}
\includegraphics[width = 0.95\columnwidth]{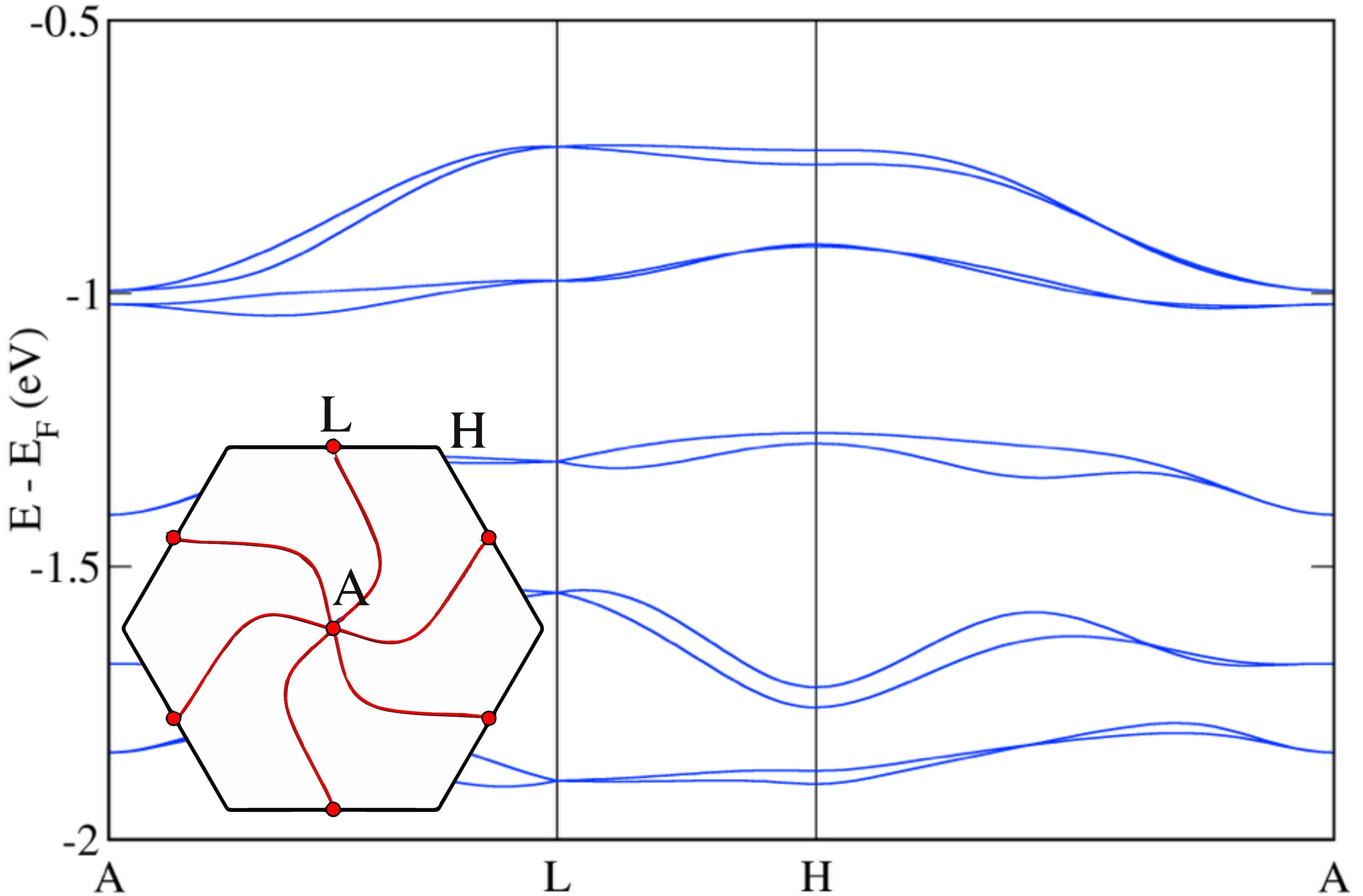}
\caption{ \label{fig_LaBr3}
First-principles band structure of LaBr$_3$ in SG $P6_3/m$ (No.~176). 
The band crossings along the A--L--H--A path are part of   
Dirac nodal lines, which are symmetry enforced by the 
off-centered symmetries $\widetilde{M}_z$ and $P$ (cf.~Sec.~\ref{mSec3}).
The inset shows the Dirac nodal lines within the $k_z= \pi$ plane formed
by the two topmost valence  bands.
All the other bands form similar nodal lines, cf.~Fig.~\ref{additional_plots_nodal_line_materials}(b).
}
\end{figure}

\paragraph{LaBr$_3$---}
An example of a hexagonal material with Dirac nodal lines
is LaBr$_3$ in SG  $P6_3/m$ (No.~176)~\cite{meyer_gerd_LaBr3_89}.
As indicated in Table~\ref{mTab1}, materials in this SG exhibit 
fourfold degenerate nodal lines within the $k_z=\pi$ plane protected
by the off-centered symmetries $\widetilde{M}_z$ and $P$.
In order to verify this, we perform first-principles band structure calculations of LaBr$_3$
to obtain the band structure shown in Figs.~\ref{fig_LaBr3} and~\ref{additional_plots_nodal_line_materials}(c).
All the bands of LaBr$_3$ are Kramers degenerate, since SG No.~176 ($P6_3/m$) contains a $PT$ symmetry which squares to $-1$.
Along the A--L--H--A path, within the $k_z = \pi$ plane, there are groups of two Kramers degenerate bands which cross each other several times.
These band crossings are part of a fourfold degenerate Dirac nodal line, whose shape resembles a star (inset of Fig.~\ref{fig_LaBr3}),
 in complete agreement with the theoretical analysis of Sec.~\ref{mSec3}. 
 These Dirac nodal lines are protected from hybridizing, since   
 the bands that cross have opposite $\widetilde{M}_z$ eigenvalues.
 Note that such star-shaped Dirac nodal lines are formed by all the bands at \emph{all} energies, since 
their existence follows from symmetry alone, independent of the energetics of the bands.
Thus, probing this insulating material below the Fermi energy (which might be possible if flakes of this layered compound are deposited on a metallic substrate) with ARPES, would reveal the star shaped 
band crossings.

 In closing we note that LiScI$_3$~\cite{lachgar1_inorganic_chemistry_91} crystallizing in SG $P\bar{6}c2$  (No.~188) is an example of a material with Weyl nodal lines within the $k_x k_z$-plane. Unfortunately, 
 the spin-orbit coupling in this material is rather weak, leading to a band splitting of only about $\sim10$~meV.
 We therefore do not discuss this material in any further detail here.

\section{Conclusions}
\label{mSec5}

In this work we have classified
all possible nonsymmorphic band degeneracies in hexagonal 
materials with time-reversal symmetry and strong spin-orbit coupling.
Our classification approach is based on representation theory of space groups and the 
algebraic relations between symmetry operators.
We find that thirteen out of the 27 hexagonal space groups (SGs)  support topological band crossings
protected by nonsymmorphic symmetries (Table~\ref{mTab1}).
Among them there are ten SGs with Weyl nodal points (Nos.~169-173 and Nos.~178-182),
two SGs with Weyl nodal lines (Nos. 188 and 190) and one SG with Dirac nodal lines (No.~176).
The stability of these band crossings is ensured by quantized topological numbers, i.e., by a
Chern number or a $\pi$-Berry phase.
We emphasize that the appearance of these band crossings is enforced by symmetry alone, i.e., they occur
in any (weakly correlated) material crystallizing in these SGs, regardless of the chemical composition.

The results of our classification are helpful for searching and designing  materials with Weyl and Dirac nodal points or nodal lines.
Using the Inorganic Crystal Structure Database (ICSD)~\cite{ICSD_link}, we have identified a number of hexagonal materials which exhibit topological band crossings (last column of Table~\ref{mTab1}).
Particularly interesting are $\gamma$-In$_2$Se$_3$ and ZrIrSn, as their band crossings are sufficiently close to the Fermi energy to be measurable.
The nontrivial topology of the band crossings in these two materials has various experimental consequences:
arc and drumhead surface states, which can be observed by angle-resolved photoemission,
and various anomalous responses to external probes, such as anomalous magnetotransport properties,
and anomalous Hall effects~\cite{burkov_review_Weyl}. These anomalous properties
may open up the possibilities for new device applications, such as spin-filter transistors~\cite{shi_wu_APL_15}, or
valleytronics applications, which utilize the valley degree of freedom  
to process information~\cite{valleytronics_review,rui_arXiv_17}.
We hope that this
will stimulate experimentalists to synthesize the reported materials and study their topological properties and responses.

In closing, we mention several interesting directions for future studies.
First, our approach can be generalized to other magnetic and nonmagnetic SGs containing screw rotations or glide mirror symmetries.
In particular, materials in SGs containing threefold or fourfold screw rotations (e.g., in trigonal systems~\cite{Trigonal_to_be_published}) 
can have multiple Weyl band crossings with accordion-like dispersions, similar to Fig.~\ref{mFig3}.
Second, it would be of interest to derive a detailed theory of the topological responses,
in particular for the Dirac nodal line materials in SG No.~176, which remain poorly understood.
Third, we expect that our analysis can be adapted to bosonic systems,  for example,
 phononic or photonic band structures, or the bands formed by elementary excitations
of quantum magnets.

\acknowledgments
The authors thank M.~Hirschmann and A.~Yaresko
for useful discussions. 
 This research was partially supported by NSF through the Princeton Center for Complex Materials, a Materials Research Science and Engineering Center DMR-1420541.
M.G.V.~was supported by IS2016-75862-P national project of the Spanish MINECO.
\mbox{C.-K.C.} acknowledges  support by Microsoft and the Laboratory for Physical Science.
J.Z.~thanks the Max-Planck-UBC-Tokyo Centre for Quantum Materials for hospitality and financial support.


\appendix

\setcounter{figure}{0}
\makeatletter 
\renewcommand{\thefigure}{A\@arabic\c@figure} 

\setcounter{equation}{0}
\makeatletter 
\renewcommand{\theequation}{A\@arabic\c@equation} 


\section{Minimum number of crossings}
\label{appendix_A}

\begin{figure}[b!]
\centering
\minipage{0.48\linewidth}
\includegraphics[width = \linewidth]{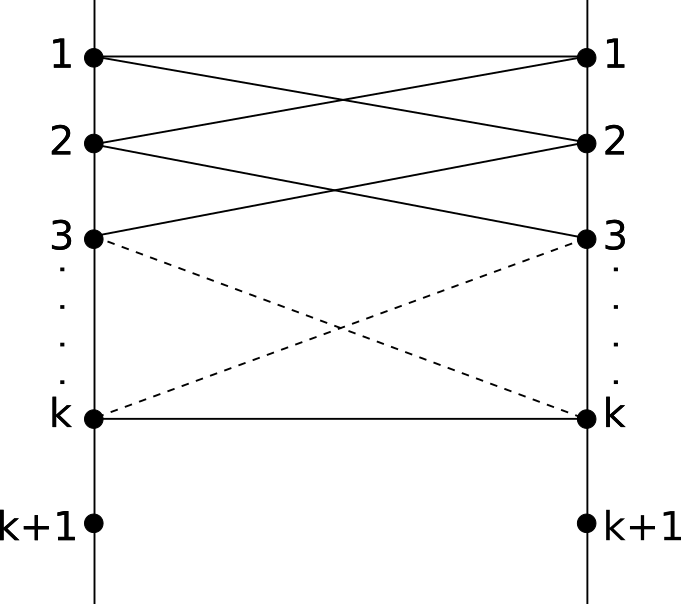}
\endminipage \hfill
\minipage{0.48\linewidth}
\includegraphics[width = \linewidth]{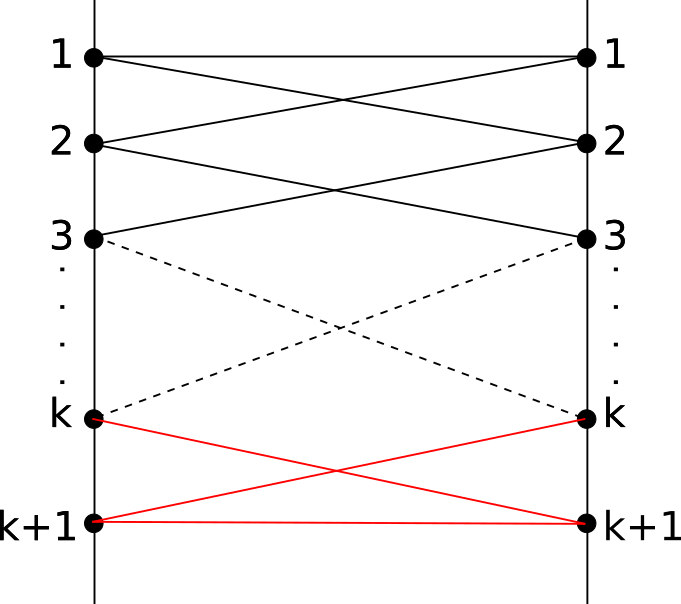}
\endminipage
\caption{ \label{bipartite_graph}
Procedure to connect one more node on each side of a bipartite graph. The added crossings are minimized by connecting the closest nodes.}
\end{figure}

In this appendix we prove that 
$2k$ bands forming a connected group along a line in between two TRIMs must 
cross at least  $k-1$ times (cf.~Sec.~\ref{mSec2A}). This problem is closely related to finding the minimum number of crossings for a bipartite graph, which in general is not a solved problem \cite{tamassia_13}. 
However, in the case of   connected groups of bands with time-reversal symmetry and spin-orbit coupling, the corresponding bipartite graphs exhibit additional constraints, such
that a proof can be constructed using mathematical induction.

We start by relating connected groups of $2k$ bands to connected bipartite graphs of order $2k$. For that purpose, we identify the energy values of the Kramer's pairs at the two TRIMs 
with the vertices (i.e., nodes) of the graph. The edges (i.e., lines) of the graph are identified with the bands along the path that connects the two TRIMs. Because this identification is one-to-one, 
the lines cannot curve outside the nodes to avoid crossings. 
Each node of the graph has degree two, because in the absence of inversion symmetry the Kramer's pairs split into two nondegenerate bands as we move away from the TRIMs.
Moreover, the graph is bipartite, since every line connects two nodes of two distinct TRIMs. The order of the graph is $2k$, because  $2k$ bands  form $2k$ Kramer's pairs at the two TRIMs.
Using mathematical induction, we will now show that the minimum number of crossings in this bipartite graph with $2k$ nodes  is $k-1$.

First we need to proof this statement for $k=1$:  The minimum number of crossings in a bipartite graph with two nodes is trivially zero,
since a double connection between two nodes never has to cross itself. For the induction step, we need to show that if the  statement holds for a graph with $2k$ nodes, it must also
hold for a graph with $2k+2$ nodes. So  let us assume that there is a graph with $2k$ nodes with crossing number $k-1$. We now wish to connect two more nodes of degree two  to the bipartite graph,
(i.e, one node at each TRIM), such that the total number of crossings is increased by as little as possible (see Fig.~\ref{bipartite_graph}).
This is achieved by (i) connecting the $(k+1)$th node at the left TRIM to the $(k+1)$th node at the right TRIM and (ii) connecting the $k$th node at the left TRIM to the $(k+1)$th node at the right TRIM, and vice versa.
This creates one additional crossing. Thus, the total number of crossings is now $k = (k+1) - 1$, which completes the proof.

\section{Representation theory}
\label{appendix_C}

To be completed.

\section{Additional band structure and surface state calculations}
\label{appendix_B}

\begin{figure}[t!]
\centering
\includegraphics[width = 0.9\linewidth]{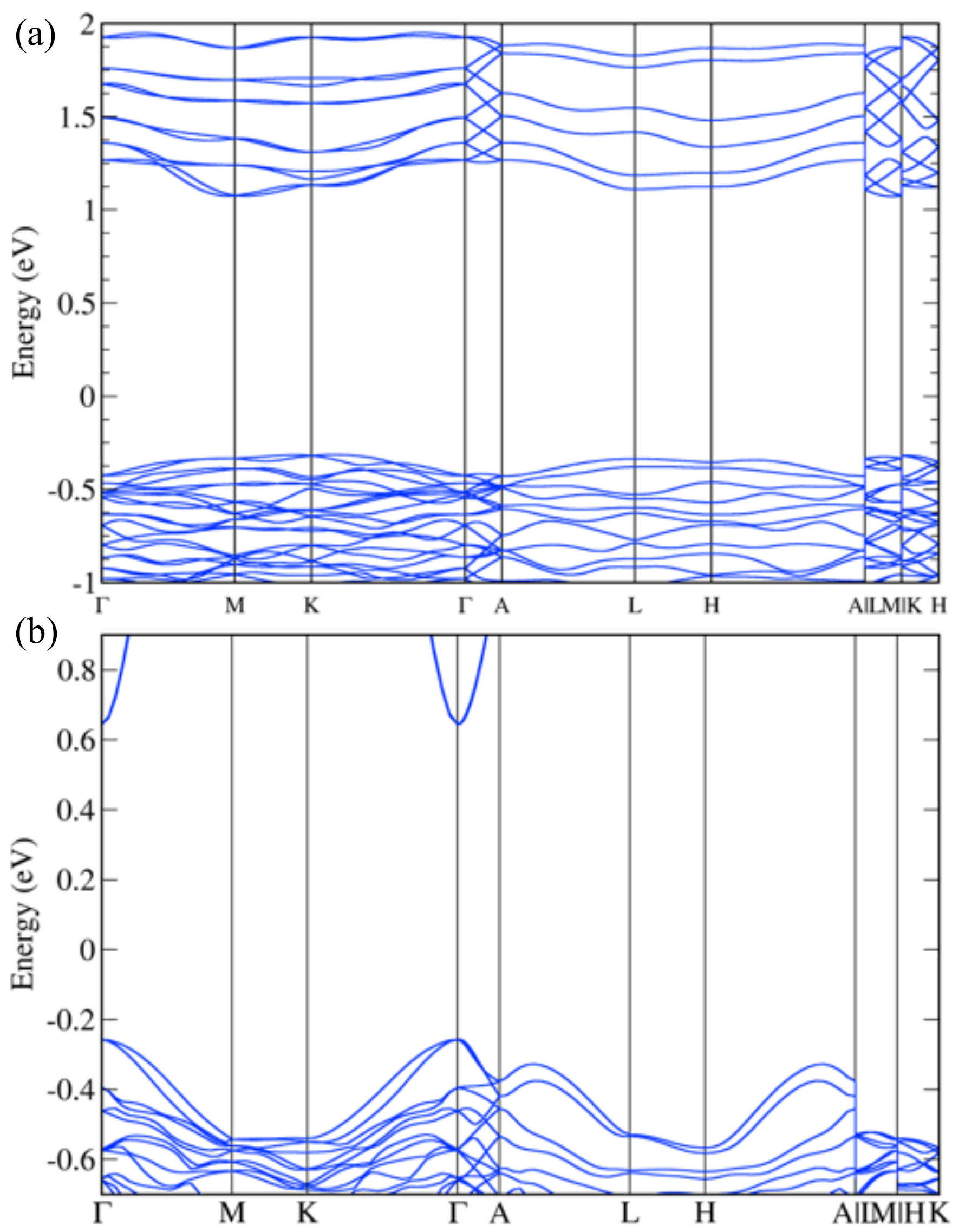}
\caption{ \label{mFig_appendix_AuF3_In2Se3}  \label{mFigA5}
DFT band structures of (a) AuF$_3$ and (b) $\gamma$-In$_2$Se$_3$, which
exhibit Weyl points along the $\Gamma$--$\Delta$--A and M--U--L lines.
Energies are measured with respect to the Fermi energy $E_{\textrm{F}}=0$.}
\end{figure}

\begin{figure}[t!]
\centering
\includegraphics[width = 0.8\linewidth]{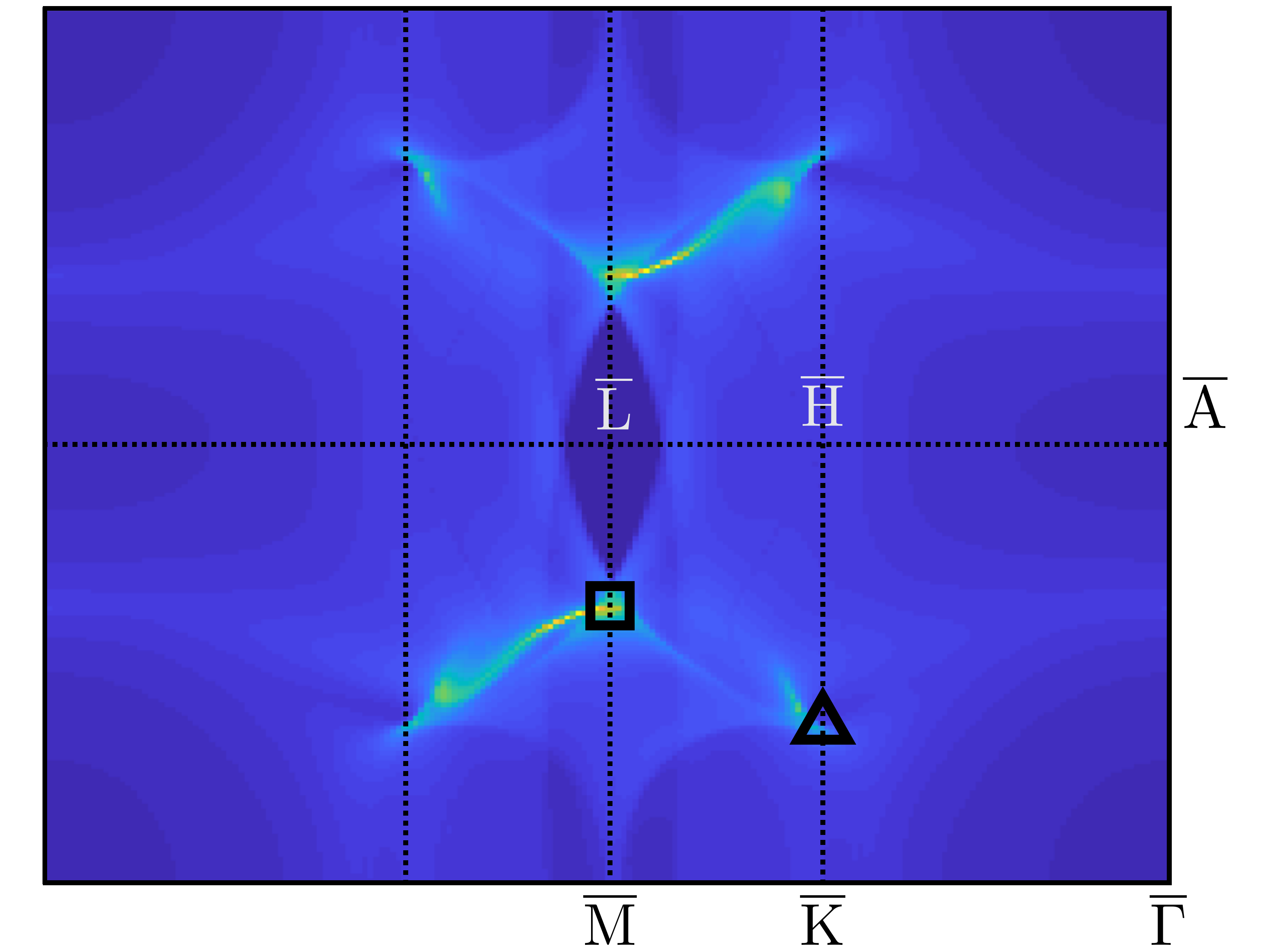}
\caption{ \label{top_surface_state_AuF3}
Momentum-resolved surface density of states for the (010) top surface of AuF$_3$ at the energy $E-E_F =1.14$~eV. The
corresponding plot for the bottom surface is shown in Fig.~\ref{fig_AuF3_surface}(a).
The arrangement of the Au and F atoms on the top and bottom terminations is presented in Figs.~\ref{fig_AuF3_surface}(b)
and \ref{fig_AuF3_surface}(c).
}
\end{figure}

\begin{figure}[ht]
\centering
\includegraphics[width = 0.9\linewidth]{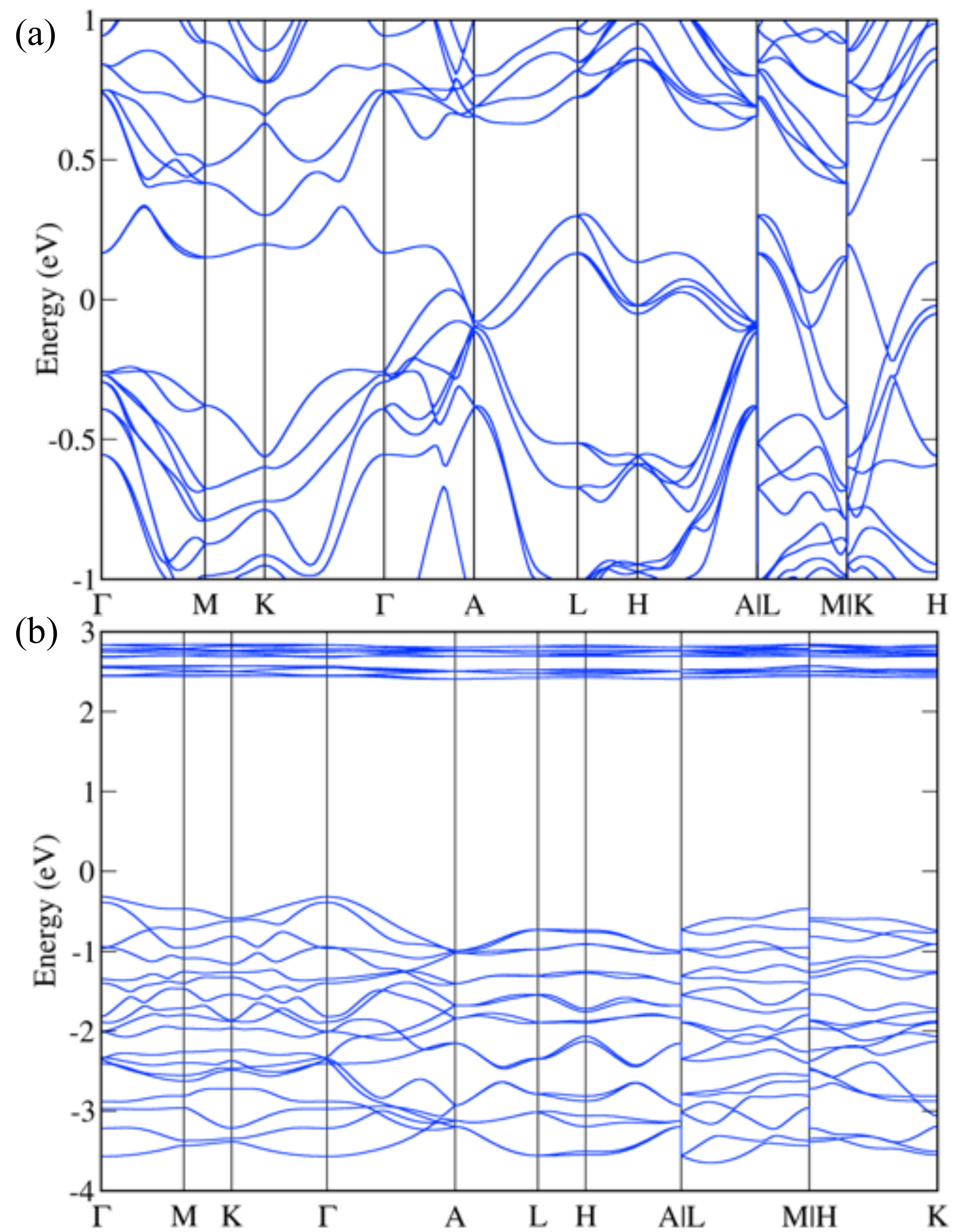}
\caption{ \label{additional_plots_nodal_line_materials} \label{mFigA6}
DFT band structures of (a) ZrIrSn and (b) LaBr$_3$, which exhibit Weyl lines and Dirac lines, respectively.
Energies are measured with respect to the Fermi energy \mbox{$E_{\textrm{F}}=0$}.}
\end{figure}

In this appendix we present additional band structure and surface state calculations for the example materials discussed in the main text. 

Figure~\ref{mFigA5} displays the full band structures of the two materials AuF$_3$ and $\gamma$-In$_2$Se$_3$, which
exhibit Weyl points. 
AuF$_3$ is insulating with a large gap of $\sim2$~eV. We observe that
Weyl points with accordion dispersions occur along $\Gamma$--$\Delta$--A both in the conduction and valence bands.
While the accordion dispersion is clearly visible above $E_{\textrm{F}}$, 
it is less clear below $E_{\textrm{F}}$, where the bands are not well separated in energy [Fig.~\ref{mFigA5}(a)]. 
Nevertheless, all valence bands form Weyl points along the $\Gamma$--$\Delta$--A and M--U--L lines, as required by symmetry.
By the bulk-boundary correspondence these Weyl points give rise to arc surface states.
Figure~\ref{top_surface_state_AuF3} shows  the  arc states at the   (010) surface of AuF$_3$ with top termination, which
connect the Weyl points formed by the  lowest two conduction bands. 
Since AuF$_3$ is insulating with a gap of $\sim2$~eV, it is very difficult to experimentally probe the Weyl points. In principle, one could try to dope
AuF$_3$ with F defects to bring the Fermi level into the valence band. But this is likely   challenging. 

The $\gamma$ phase of In$_2$Se$_3$ is also insulating, but with a smaller gap of $\sim0.9$~eV, see Fig.~\ref{mFigA5}(b). 
Here, it might be possible to measure the Weyl points along the $\Gamma$--$\Delta$--A line using photoemission, as they are only
about 300-400~meV below the Fermi energy. It might also be possible to observe signatures of the
Fermi arc states  using Fourier-transform scanning tunneling spectroscopy.

In Fig.~\ref{mFigA6} we present the full band structures of the nodal line materials ZrIrSn and LaBr$_3$, which exhibit Weyl and Dirac nodal lines, respectively. 
ZrIrSn contains two Weyl nodal lines within the $k_x = \pi$ plane  
that are only about 100 meV away from the Fermi energy (Figs.~\ref{mFigA6}(a) and~\ref{fig_IrSnZr}).
These Weyl nodal lines give rise to drumhead states on the (100) and (010) surfaces.
Because of spin-orbit coupling, these drumhead surface states posses an
intricate helical spin texture, with the spin and momentum directions locked to each other.
It should be possible to measure this spin texture using, e.g., spin-resolved photoemission or
spin-resolved scanning tunneling spectroscopy.
Furthermore, in slightly doped ZrIrSn samples the Weyl nodal lines could 
be accessed in transport experiments.
We expect that this material will show anomalous magnetoelectric responses and anomalous Hall effects, since the
bands carry a nonzero Berry curvature.
 
The Dirac nodal lines of LaBr$_3$, which are located within the $k_z = \pi$ plane, are difficult to observe, as this material is insulating
with a gap of about $\sim 2.5$~eV [Fig.~\ref{additional_plots_nodal_line_materials}(b)].
It might be possible to probe the Dirac nodal lines below the Fermi energy in exfoliated flakes of LaBr$_3$ deposited on a metallic substrate.
However, this is likely challenging as the material is air sensitive. 
 Therefore, the search for suitable Dirac nodal-line materials in SG 176 with band crossing at the Fermi energy
remains as a goal for future work.


\bibliographystyle{apsrev4-1}
\bibliography{hexagonals_literature}

\appendix

\end{document}